\documentclass[11pt]{article}
\PassOptionsToPackage{authoryear, sort}{natbib}
\usepackage[preprint]{neurips_2020}

\usepackage{xcolor}
\usepackage{mathtools}
\usepackage[utf8]{inputenc} % allow utf-8 input
\usepackage[T1]{fontenc}    % use 8-bit T1 fonts
\usepackage[hidelinks]{hyperref}       % hyperlinks
\usepackage{url}            % simple URL typesetting
\usepackage{booktabs}       % professional-quality tables
\usepackage{amsmath,amsfonts,amssymb}
\usepackage{nicefrac}       % compact symbols for 1/2, etc.
\usepackage{microtype}
\usepackage[labelfont=bf]{caption}
\usepackage{indentfirst}
\usepackage{etoolbox}
\apptocmd{\sloppy}{\hbadness 10000\relax}{}{}

\usepackage{lineno}
\DeclareMathOperator{\sgn}{sgn}
\newcommand{\ie}{{\it i.e.}}
\renewcommand{\d}{\mathrm{d}}

\newcommand{\tens}[1]{{\boldsymbol{#1}}}
\newcommand*{\pd}[3][]{{\frac{\partial^{#1} #2}{\partial #3}}}
\newcommand{\bracs}[1]{\left({#1}\right)}
\newcommand{\bracm}[1]{\left[{#1}\right]}
\newcommand{\abs}[1]{\left|{#1}\right|}
\newcommand*{\dydx}[3][]{\frac{\mathrm{d}^{#1} {#2}}{\mathrm{d} {#3}}}
\title{Fluid flow through anisotropic and deformable double porosity media with ultra-low matrix permeability: A continuum framework}

\author{Qi Zhang\thanks{Corresponding author} \\
  Department of Civil and Environmental Engineering\\
  Stanford University \\
  Stanford, CA 94305, USA \\
  \texttt{qzhang94@stanford.edu} \\
  \And
  Xia Yan \\
  School of Petroleum Engineering \\
  China University of Petroleum (East China) \\
  Qingdao, 266580, China \\
  \texttt{jsyanxia1989@163.com} \\
  \And
  Jianli Shao \\
  State Key Laboratory of Mining Disaster Prevention and Control \\
  Shandong University of Science and Technology \\
  Qingdao, 266590, China \\
  \texttt{jianli.shao@sdust.edu.cn} \\
  \texttt{shaojianli5020@163.com}
}

\begin{document}
\maketitle

\begin{abstract}
Fractured porous media or double porosity media are common in nature. At the same time, accurate modeling remains a significant challenge due to bi-modal pore size distribution, anisotropy, multi-field coupling, and various flow patterns. This study aims to formulate a comprehensive coupled continuum framework that could adequately consider these critical characteristics. In our framework, fluid flow in the micro-fracture network is modeled with the generalized Darcy's law, in which the equivalent fracture permeability is upscaled from the detailed geological characterizations. The liquid in the much less permeable matrix follows a low-velocity non-Darcy flow characterized by threshold values and non-linearity. The fluid mass transfer is assumed to be a function of the shape factor, pressure difference, and (variable) interface permeability. The solid deformation relies on a thermodynamically consistent effective stress derived from the energy balance equation, and it is modeled following anisotropic poroelastic theory. The discussion revolves around generic double porosity media. Model applications reveal the capability of our framework to capture the crucial roles of coupling, poroelastic coefficients, anisotropy, and ultra-low matrix permeability in dictating the pressure and displacement fields.
\end{abstract}
\textbf{Keywords.} Double porosity; Geomechanics; Upscaling; Anisotropy; Non-Darcy parameter; Consolidation

\section{Introduction}
\label{intro}
In our natural environment, the real reservoirs tend to be very heterogeneous in both porosity and permeability characteristics due to the existence of porous constituents at various length scales \citep{Ashworth2019}. The accurate modeling of real reservoirs remains a significant challenge. Instead, people always idealize the actual reservoir as an aggregate of different geological regions (e.g., host rock, fracture, fault, compaction band, and so on), among which the double porosity model \citep{Barenblatt1960,Warren1963,Wilson1982} is widely adopted in engineering practice. In double porosity model, there are transport porosity which appears in the form of micro-fractures/fissures, and storage porosity which appears in the form of matrix pores/nanopores. It must be noted that the macro-fractures such us hydraulic fractures constitute another (much larger) porosity scale \citep{Zhang2018}, which is not considered in the typical double porosity model, thus the fractures appeared in this paper should be understood as the micro-fractures or natural fractures. Generally speaking, there are two classes of methods used for modeling double porosity media: discrete (explicit) method and continuum (implicit) method \citep{Ashworth2019}. In this paper, we focus on the continuum method, while at the same time, we try to find its relation with the discrete descriptions of the micro-fractures.

Darcy's law is the most fundamental equation to describe fluid flow \citep{Shao2020a,Shao2020b}. However, for unconventional and tight reservoirs with ultra-low matrix permeability (assuming isotropic), Darcy's law could overestimate the flow rate of liquid due to the interaction between fluid particles and the solid pore wall \citep{Xiong2017,Wang2017,Dmitriyev2001}. The result of this interaction is the formation of a boundary layer on which the liquid exhibits higher viscosity \citep{Wang2017}. It has been argued in the literature that when the magnitude of gradient is extremely small, say, lower than a scalar called threshold pressure gradient (TPG) \citep{Hao2008}, the boundary layer could prevent the fluid from flowing, and above this TPG, the flow curve in each direction shows a certain level of non-linearity \citep{Wang2011}, followed by a straight line \citep{Li2016}. In other words, Darcy's law should be corrected for the effect of the TPG. However, there also exist some opposite opinions about TPG, which asserted such threshold does not exist or it is a misinterpretation of experimental data \citep{Wang2017}. Instead, \citet{Wang2017} chose an alternative nonlinear model.

In addition to the extensive investigations on the flow problem, the tightly coupled hydro-mechanical behavior is central to the performance of many subsurface systems and is critical for assessing environmental impacts \citep{Castelletto2015}. This strong coupling is always modeled as a two-way coupled process, which is described by the well-established poromechanical theory for conventional single porosity media \citep{Zhang2020c}. For double porosity media, \citet{Wilson1982} made the first attempt to consider solid deformation in double porosity media by introducing new governing equations as well as new material properties. Over the last 30 years, efforts along this line have resulted in many different modeling approaches that are individually developed \citep{Bai1993,Mehrabian2014,Berryman1995,Khalili2003,Ghafouri1996}. Almost all the discrepancies come from the actual modeling of porosity change and this remains, to the best of current authors' knowledge, an open question. Furthermore, these previous formulations have assumed isotropy in both deformation and fluid flow, while it is well-known that many geologic materials have exhibited anisotropy in either or both deformation and fluid flow responses. For a double porosity medium, however, the effect of anisotropy has not been clearly elucidated in light of the limitations imposed by current laboratory testing procedures.

\textcolor{black}{This paper aims to address the above-mentioned knowledge gap by developing a mathematically consistent framework for fluid flow through anisotropic and deformable double porosity media with ultra-low matrix permeability. A novel feature of the mathematical formulation entails the use of the newly proposed constitutive laws for $\d \phi_1/ \d t$ and $\d \phi_2/ \d t$ in combination with mixture theory to arrive at the governing fluid flow and solid deformation equations. Therefore, the fundamental origins of model parameters are clearly established. The mathematical model is innovative because it is still consistent with some previous isotropic models, but our model gives more freedom to investigate new hydro-mechanical patterns (see Section~\ref{wilson} for the consolidation with double porosity) and incorporate new constitutive laws or coupling fields in the future. It is the first time, to the authors' knowledge, that these new formulas and interpretations are presented within the context of poromechanics. Another novel contribution of this paper is an upscaling approach based on the volume integral (with an illustrative application/example), which is useful in shedding light onto the physical meaning of the equivalent fracture permeability.}

\section{Mathematical formulations}
\subsection{Fluid flow model}
\begin{figure}[htb]
	\begin{center}
	\begin{tabular}{c}
		\includegraphics[width=0.5\textwidth]{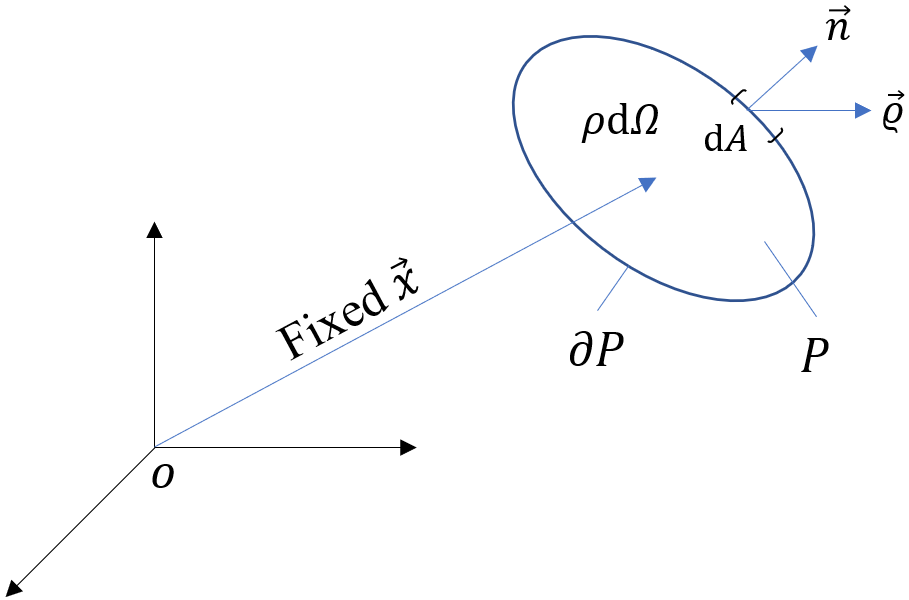}
	\end{tabular}
	\end{center}
	\caption
	{\label{Fig:01}
An arbitrary control volume $P$ within the computational domain where $\rho$ is a density term, $\vec{\varrho}$ is the corresponding flux term of $\rho$, and $\vec{n}$ is the unit outward normal vector of the infinitesimal area $\d A$. The $\rho$ and $\vec{\varrho}$ are specified in Figure~\ref{Fig:02}.}
\end{figure}

\begin{figure}[htb]
	\begin{center}
	\begin{tabular}{c}
		\includegraphics[width=0.4\textwidth]{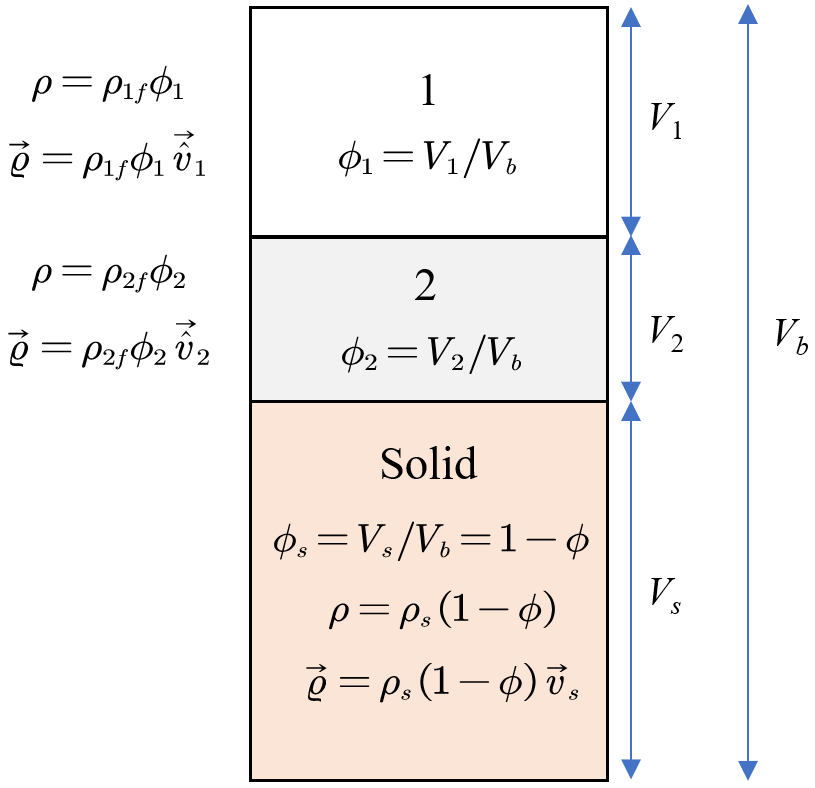}
	\end{tabular}
	\end{center}
	\caption
	{\label{Fig:02}
Phase diagram of the double porosity media.}
\end{figure}

The fluid flow model is established based on the arbitrary control volume of Figure~\ref{Fig:01}. For double porosity media shown in Figure~\ref{Fig:02}, by specifying $\rho$ and $\vec{\varrho}$, we could get three mass conservation equations in their original forms \citep{Coussy2003,Kim2010,Khalili2003}:
\begin{equation}
	\label{root_eq1}
	\frac{\partial}{\partial t}\bracm{\rho_s \bracs{1 - \phi}} + \nabla\cdot \bracm{\rho_s \bracs{1 - \phi} \vec{v}_s} = 0\,,
\end{equation}
\begin{equation}
	\label{root_eq2}
	\frac{\partial}{\partial t}\bracm{\rho_{1f} \phi_1} + \nabla\cdot \bracm{\rho_{1f} \phi_1 \vec{\hat{v}}_1} = \rho_{1f} c_1\,,
\end{equation}
\begin{equation}
	\label{root_eq3}
	\frac{\partial}{\partial t}\bracm{\rho_{2f} \phi_2} + \nabla\cdot \bracm{\rho_{2f} \phi_2 \vec{\hat{v}}_2} = \rho_{2f} c_2\,,
\end{equation}
\textcolor{black}{where $\phi_1 = V_1/V_b$ is the Eulerian porosity of the matrix pores or nanopores, $\phi_2 = V_2/V_b$ is the Eulerian porosity of the micro-fracture network, $\phi = \phi_1 + \phi_2$ is the total porosity, $\phi_s = 1 - \phi = V_s/V_b$ is the volume fraction of the solid,} $\rho_s$ is the solid grain density, $\rho_{1f}$ and $\rho_{2f}$ are the fluid densities, $\vec{v}_s$ is the velocity of the solid skeleton, $\vec{\hat{v}}_1$ and $\vec{\hat{v}}_2$ are the interstitial velocities of the fluid particles, $c_1$ and $c_2$ represent the source/sink terms. In this paper, we assume mass exchange could only happen between fluids, which implies $c_1 + c_2 = 0$.

However, we cannot solve above equations directly, and further transformations are necessary and also important. To begin with, we need to introduce the material time derivatives whose definition are given as follows:
\begin{equation}
	\label{mtd_s}
	\dydx[s]{\bracs{\cdot}}{t} = \pd{\bracs{\cdot}}{t} + \nabla\bracs{\cdot}\cdot\vec{v}_s\,,
\end{equation}
\begin{equation}
	\label{mtd_1}
	\dydx[1f]{\bracs{\cdot}}{t} = \pd{\bracs{\cdot}}{t} + \nabla\bracs{\cdot}\cdot\vec{\hat{v}}_1\,,
\end{equation}
\begin{equation}
	\label{mtd_2}
	\dydx[2f]{\bracs{\cdot}}{t} = \pd{\bracs{\cdot}}{t} + \nabla\bracs{\cdot}\cdot\vec{\hat{v}}_2\,.
\end{equation}
In the following text, since we prefer material time derivative following the solid phase motion whenever possible, we could drop the superscript $s$ in $\mathrm{d}^s \bracs{\cdot}/\mathrm{d} t$. For Eq.~\eqref{root_eq2} and Eq.~\eqref{root_eq3}, the following processes are basically the same except for the change of the subscript. Thus we deal with Eq.~\eqref{root_eq2} in detail and provide the final form of Eq.~\eqref{root_eq3} at the end of our derivation. By expanding derivatives in Eq.~\eqref{root_eq2}, we get:
\begin{equation}
	\label{expand_derivative}
	\rho_{1f} \pd{\phi_1}{t} + \phi_1\pd{\rho_{1f}}{t} + \phi_1\vec{\hat{v}}_1 \cdot \nabla \rho_{1f} + \rho_{1f}\nabla\cdot\bracm{\phi_1 \vec{\hat{v}}_1} = \rho_{1f} c_1\,.
\end{equation}
In Eq.~\eqref{expand_derivative}, after extraction of $\phi_1$, the combination of the second and the third terms on the left-hand side is exactly the material time derivative shown in Eq.~\eqref{mtd_1}. Adopting this definition and further dividing $\rho_{1f}$ on both sides, we obtain:
\begin{equation}
	\label{eq:9}
	\pd{\phi_1}{t} + \frac{\phi_1}{\rho_{1f}}\dydx[1f]{\rho_{1f}}{t} + \nabla\cdot\bracm{\phi_1 \vec{\hat{v}}_1} = c_1\,.
\end{equation}
Now we need to define superficial velocity $\vec{q}_1$ in our context as:
\begin{equation}
	\label{darcyvel}
	\vec{q}_1 = \phi_1\bracs{\vec{\hat{v}}_1 - \vec{v}_s}\,.
\end{equation}
Then we can rewrite Eq.~\eqref{eq:9} using $\vec{q}_1$ and we have:
\begin{equation}
	\label{eq:11}
	\pd{\phi_1}{t} + \frac{\phi_1}{\rho_{1f}}\dydx[1f]{\rho_{1f}}{t} + \nabla\cdot\vec{q}_1 + \nabla\cdot\bracm{\phi_1 \vec{v}_s} = c_1\,.
\end{equation}
By further making use of the following relation:
\begin{equation}
	\label{eq:12}
	\pd{\phi_1}{t} + \nabla\cdot\bracm{\phi_1 \vec{v}_s} = \pd{\phi_1}{t} + \vec{v}_s\cdot\nabla \phi_1 + \phi_1 \nabla\cdot\vec{v}_s = \dydx{\phi_1}{t} + \phi_1 \nabla\cdot\vec{v}_s\,,
\end{equation}
we would obtain:
\begin{equation}
	\label{eq:13}
	\frac{\phi_1}{\rho_{1f}}\dydx[1f]{\rho_{1f}}{t} + \nabla\cdot\vec{q}_1 + \dydx{\phi_1}{t} + \phi_1 \nabla\cdot\vec{v}_s = c_1\,.
\end{equation}
In this paper, we assume the fluid density $\rho_{1f}$ is a function of the fluid pressure $p_1$ given as:
\begin{equation}
	\label{PVT-1}
	\rho_{1f} = \rho_{f}\exp\bracs{\frac{p_1 - p^0}{K_{1f}}}\,,
\end{equation}
where $\rho_f$ is the reference fluid density, $p^0$ is the reference pressure, and $K_{1f}$ is the fluid bulk modulus. Combining Eq.~\eqref{PVT-1} and Eq.~\eqref{eq:13} gives:
\begin{equation}
	\label{root-transformed_eq2}
	\frac{\phi_1}{K_{1f}}\dydx[1f]{p_1}{t} + \nabla\cdot\vec{q}_1 + \dydx{\phi_1}{t} + \phi_1 \nabla\cdot\vec{v}_s = c_1\,.
\end{equation}
Exactly the same logic also applies to Eq.~\eqref{root_eq3}. As a result, we get:
\begin{equation}
	\label{root-transformed_eq3}
	\frac{\phi_2}{K_{2f}}\dydx[2f]{p_2}{t} + \nabla\cdot\vec{q}_2 + \dydx{\phi_2}{t} + \phi_2 \nabla\cdot\vec{v}_s = c_2\,.
\end{equation}

Until now, except for the assumption of the basic double porosity model which admits two overlapping continua, no additional assumption has been made in the preceding derivations to Eqs.~\eqref{root-transformed_eq2}\eqref{root-transformed_eq3}. In other words, we can regard Eqs.~\eqref{root-transformed_eq2}\eqref{root-transformed_eq3} as the starting point to introduce all kinds of specific constitutive laws, among them the modeling of $\mathrm{d}\phi_1/\mathrm{d}t$ and $\mathrm{d}\phi_2/\mathrm{d}t$ is still an open question, and that is why we have so many different modeling approaches mentioned in Section~\ref{intro}. In this paper, we try to move one step further by proposing a new expression inspired by \citet{Cheng1997,Cheng2016,Ashworth2019,Kim2010} that could incorporate all the other existing modeling approaches. To motivate our expression, let us go back to the single porosity media. In single porosity media, we have the following relation for the evolution of the \textcolor{black}{Eulerian porosity $\phi$} \citep{Coussy2003}:
\begin{equation}
	\label{phi-evol}
		\dydx{\phi}{t} = \frac{1}{N}\dydx{p}{t} + \bracs{\tens{\alpha} - \phi \tens{1}}:{\dydx[]{\tens{\epsilon}}{t}}\,,
\end{equation}
where $1/N$ is the inverse of Biot's tangent modulus \citep{Coussy2003}, $p$ is the fluid pressure, $\tens{\alpha}$ is the Biot's symmetric tangent tensor \citep{Coussy2003}, $\tens{1}$ is the second-order identity tensor, $\tens{\epsilon} = \bracs{\nabla \vec{u} + \nabla^T \vec{u}}/2$ is the infinitesimal strain tensor, and $\vec{u}$ is the solid displacement vector ($\vec{v}_s = \d \vec{u}/ \d t$). Motivated by Eq.~\eqref{phi-evol} and given we now have two pressure fields and two porous regions ($V_1$ and $V_2$), we can propose the following expression to model $\mathrm{d}\phi_1/\mathrm{d}t$ and $\mathrm{d}\phi_2/\mathrm{d}t$:
\begin{equation}
	\label{phi1phi2-evol}
	\begin{Bmatrix}
		{\d \phi_1}/{\d t}\\
		{\d \phi_2}/{\d t}
	\end{Bmatrix} 
	= \begin{bmatrix}
		N_{11} & N_{12} \\
		N_{12} & N_{22}
	\end{bmatrix}
	\begin{Bmatrix}
		{\d p_1}/{\d t} \\
		{\d p_2}/{\d t}
	\end{Bmatrix} + \begin{bmatrix}
		\bracs{\tens{\alpha}_1 - \phi_1 \tens{1}}:{\d \tens{\epsilon}}/{\d t} \\
		\bracs{\tens{\alpha}_2 - \phi_2 \tens{1}}:{\d \tens{\epsilon}}/{\d t}
	\end{bmatrix}\,,
\end{equation}
where the coefficient $1/N$ in Eq.~\eqref{phi-evol} becomes a $2 \times 2$ pressure coupling matrix (symmetric) $N \in \mathbb{R}^{2\times 2}$, and we have two Biot tensors $\tens{\alpha}_1$ and $\tens{\alpha}_2$. A back substitution of Eq.~\eqref{phi1phi2-evol} into Eqs.~\eqref{root-transformed_eq2}\eqref{root-transformed_eq3} leads to (note $\nabla\cdot \vec{v}_s = \tens{1}:{\d \tens{\epsilon}}/{\d t}$):
\begin{equation}
\label{eq:18}
    \frac{\phi_1}{K_{1f}}\dydx[1f]{p_1}{t} + N_{11}\dydx{p_1}{t} + N_{12}\dydx{p_2}{t} + \tens{\alpha}_1:\dydx{\tens{\epsilon}}{t} + \nabla\cdot\vec{q}_1 = c_1\,,
\end{equation}
\begin{equation}
\label{eq:19}
    \frac{\phi_2}{K_{2f}}\dydx[2f]{p_2}{t} + N_{12}\dydx{p_1}{t} + N_{22}\dydx{p_2}{t} + \tens{\alpha}_2:\dydx{\tens{\epsilon}}{t} + \nabla\cdot\vec{q}_2 = c_2\,.
\end{equation}
In the actual model computation, the material time derivatives are always approximated using the partial time derivatives, and all the coefficients are evaluated using their initial values, which give us:
\begin{equation}
\label{eq:20}
    A_{11}\pd{p_1}{t} + A_{12}\pd{p_2}{t} + \tens{\alpha}_1:\pd{\tens{\epsilon}}{t} + \nabla\cdot\vec{q}_1 = c_1\,,
\end{equation}
\begin{equation}
\label{eq:21}
    A_{12}\pd{p_1}{t} + A_{22}\pd{p_2}{t} + \tens{\alpha}_2:\pd{\tens{\epsilon}}{t} + \nabla\cdot\vec{q}_2 = c_2\,,
\end{equation}
where $A_{11} = N_{11} + \phi_1/K_{1f}$, $A_{12} = N_{12}$, and $A_{22} = N_{22} + \phi_2/K_{2f}$. Besides $\mathrm{d}\phi_1/\mathrm{d}t$ and $\mathrm{d}\phi_2/\mathrm{d}t$, we still need to provide expressions for $\vec{q}_1$, $\vec{q}_2$, $c_1$, and $c_2$ to make Eqs.~\eqref{eq:20}\eqref{eq:21} solvable.

First of all, for $\vec{q}_1$, we assume it follows the low-velocity non-Darcy flow of liquid that can be expressed as a nonlinear function of $\vec{\varphi} = \nabla p_1 - \rho_{1f} \vec{g}$:
\begin{equation}
	\label{nondarcy-q1}
	\bracs{\vec{q}_1}_j = \begin{dcases}
		0 & \abs{\varphi_j} < \lambda_{\min}\,, \\
	-\frac{k_1 \bracs{\abs{\varphi_j} -\lambda_{\min}}^{\,\xi}}{\mu_{f} \, \xi \, \delta^{\,\xi-1}}  \sgn\bracs{\varphi_j}  & \lambda_{\min} \leq \abs{\varphi_j}  \leq \lambda_{\max}\,, \\
	-\frac{k_1}{\mu_{f}}\bracs{\abs{\varphi_j} - \lambda_{\min} - \frac{\xi-1}{\xi}\delta} \sgn\bracs{\varphi_j} & \abs{\varphi_j} > \lambda_{\max}\,, \\
	\end{dcases}
\end{equation}
where $\vec{g}$ is the gravity acceleration vector, $j = 1, \cdots, n_{\rm dim}$, $n_{\rm dim}$ is the space dimension, $k_1$ is the Darcy permeability of the matrix, $\mu_{f}$ is the fluid viscosity, $\xi \geq 1$ is the exponent parameter \citep{Zhang2020b}, $\delta = \lambda_{\max} - \lambda_{\min}$, $\lambda_{\min}$ has the the physical meaning of threshold gradient and $\lambda_{\max}$ has the physical meaning of critical gradient \citep{Zhang2020b}, the pseudo gradient \citep{Li2016} can be calculated analytically as $\lambda_{\min} + \delta \bracs{\xi-1}/{\xi}$. In this paper, we assume $\xi$, $\lambda_{\min}$, and $\lambda_{\max}$ are constants. From Eq.~\eqref{nondarcy-q1}, we can see that this non-Darcy flow model is a combination of a no flow part, a nonlinear flow part, and a linear flow part, separated by $\lambda_{\min}$ and $\lambda_{\max}$, which describes exactly the type curve of non-Darcy flow \citep{Zhao2020}. In addition, if we set $\xi = 1$ and $\lambda_{\min} = 0$, isotropic Darcy's law is automatically recovered. Thus Eq.~\eqref{nondarcy-q1} is more general than similar equations in \citet{Li2016,Hansbo1997}. Secondly, for $\vec{q}_2$, it is given as:
\begin{equation}
\vec{q}_2 = -\frac{\tens{k}_2}{\mu_f}\cdot\bracs{\nabla{p_2} - \rho_{2f}\vec{g}}\,,
\end{equation}
where $\tens{k}_2$ is the equivalent fracture permeability, and its calculation will be elaborated in Section~\ref{upscale}. Finally, the $c_1$ and $c_2$ are calculated as:
\begin{equation}
\label{c_1c_2}
    c_1 = - c_2 = \gamma \bracs{p_2 - p_1} = \frac{\sigma_{\rm sh} \bar{k}}{\mu_f}\bracs{p_2 - p_1}\,,
\end{equation}
where $\sigma_{\rm sh}$ is the shape factor, $\bar{k}$ is the interface permeability, $\gamma = \sigma_{\rm sh} \bar{k}/\mu_f$ is the leakage coefficient. From \citet{Khalili1999}, we know $\bar{k}$ is closely related to the apparent permeability $\partial \vec{q}_1/ \partial \vec{\varphi}$ of the matrix, thus we propose the following form of $\bar{k}$:
\begin{equation}
	\label{nondarcy-leakage}
	\bar{k} = \begin{dcases}
	\bar{k}_{\min} & \varphi_{\max} < \lambda_{\min}\,, \\	
	\frac{\bar{k}_{\max} + \bar{k}_{\min}}{2} + \frac{\bar{k}_{\min} - \bar{k}_{\max}}{2}\cos\bracm{\pi\bracs{\frac{\varphi_{\max} - \lambda_{\min}}{\delta}}^{\xi}}  & \lambda_{\min} \leq \varphi_{\max} \leq \lambda_{\max}\,, \\
	\bar{k}_{\max} & \varphi_{\max} > \lambda_{\max}\,, \\
	\end{dcases}
\end{equation}
where $\bar{k}_{\max}$ and $\bar{k}_{\min}$ are the maximum and minimum interface permeabilities, respectively, and
\begin{equation}
\varphi_{\max} = \max_{j \in \{1, \cdots, n_{\rm dim}\}} \abs{\varphi_j}\,.
\end{equation}
Note when $\bar{k}_{\max} = \bar{k}_{\min}$, the interface permeability $\bar{k}$ becomes a trivial constant as in the Warren and Root model \citep{Warren1963}.

\subsection{Solid deformation model}
The solid deformation model is based on the linear momentum balance equation of the whole medium, which is given as:
\begin{equation}
\label{linearmom}
    \nabla\cdot\tens{\sigma} + \rho \vec{g} = \vec{0}\,,
\end{equation}
where $\tens{\sigma}$ is the total stress tensor, $\rho = \rho_s \bracs{1 - \phi} + \rho_{1f} \phi_1 + \rho_{2f} \phi_2$ is the bulk density. However, the total stress tensor $\tens{\sigma}$ does not solely depend on the strain tensor $\tens{\epsilon}$, so determination of the mathematical form of the effective stress tensor $\bar{\tens{\sigma}}$ (solely depends on $\tens{\epsilon}$) is crucial for constitutive modeling in poromechanics. Here we adopt a typical energy approach to derive the energy-conjugate pair related to solid deformation. The approach starts by writing the internal energy rate $\d \mathcal{E}/ \d t$ as:
\begin{equation}
\label{eq:28}
    \dydx{\mathcal{E}}{t} =  \tens{\sigma}:\dydx{\tens{\epsilon}}{t} -\phi_1 p_1 \nabla \cdot \bracs{\vec{\hat{v}}_1 - \vec{v}_s} - \phi_2 p_2 \nabla \cdot \bracs{\vec{\hat{v}}_2 - \vec{v}_s} + \mathcal{F}\,,
\end{equation}
where $\mathcal{F}$ collects other flow and mass transfer terms that might influence the internal energy $\mathcal{E}$, but they are not our focuses here. From the definition of $\vec{q}_1$ and $\vec{q}_2$, we know:
\begin{equation}
	\label{eq:29}
	\nabla \cdot \vec{q_1} = \nabla \cdot \bracm{\phi_1\bracs{\vec{\hat{v}}_1 - \vec{v}_s}} = \phi_1 \nabla \cdot \bracs{\vec{\hat{v}}_1 - \vec{v}_s} + \bracs{\vec{\hat{v}}_1 - \vec{v}_s} \cdot \nabla \phi_1\,,
\end{equation}
\begin{equation}
	\label{eq:30}
	\nabla \cdot \vec{q_2} = \nabla \cdot \bracm{\phi_2\bracs{\vec{\hat{v}}_2 - \vec{v}_s}} = \phi_2 \nabla \cdot \bracs{\vec{\hat{v}}_2 - \vec{v}_s} + \bracs{\vec{\hat{v}}_2 - \vec{v}_s} \cdot \nabla \phi_2\,.
\end{equation}
At the same time, $\nabla\cdot\vec{q}_1$ and $\nabla\cdot\vec{q}_2$ also appear in Eq.~\eqref{eq:18} and Eq.~\eqref{eq:19}, respectively. As a result, we can use Eqs.~\eqref{eq:18}\eqref{eq:19}\eqref{eq:29}\eqref{eq:30} to rewrite Eq.~\eqref{eq:28} as:
\begin{equation}
		\label{done}
		\dydx{\mathcal{E}}{t} = \bracs{\tens{\sigma} + \tens{\alpha}_1 p_1 + \tens{\alpha}_2 p_2}:\dydx{\tens{\epsilon}}{t} + \bar{\mathcal{F}} \,,
\end{equation}
where $\bar{\mathcal{F}}$ collects the remaining terms. From Eq.~\eqref{done}, the $\tens{\sigma} + \tens{\alpha}_1 p_1 + \tens{\alpha}_2 p_2$ and $\tens{\epsilon}$ are identified as the energy-conjugate pair related to solid deformation, which means the effective stress $\bar{\tens{\sigma}}$ should be defined as:
\begin{equation}
    \bar{\tens{\sigma}} = \tens{\sigma} + \tens{\alpha}_1 p_1 + \tens{\alpha}_2 p_2\,.
\end{equation}
For linear elasticity, we have:
\begin{equation}
    \bar{\tens{\sigma}} = \mathbb{C}^e:\tens{\epsilon}\,,
\end{equation}
where $\mathbb{C}^e$ is a rank-four tensor (with major and minor symmetries) characterizing the elastic isotropy or anisotropy of the porous material. In this paper, the $\mathbb{C}^e$ is characterized by five constants $E_h$, $E_v$, $\nu_{vh}$, $\nu_{hh}$, and $G_{vh}$ for a horizontally layered material, which is also known as the VTI elasticity \citep{Lora2016}. For detailed Voigt and tensorial forms of $\mathbb{C}^e$ or $\bracs{\mathbb{C}^e}^{-1}$, please refer to \citet{Zhang2020a,Lora2016}.

\subsection{Summary}
\begin{figure}[htb]
	\begin{center}
	\begin{tabular}{c}
		\includegraphics[width=0.8\textwidth]{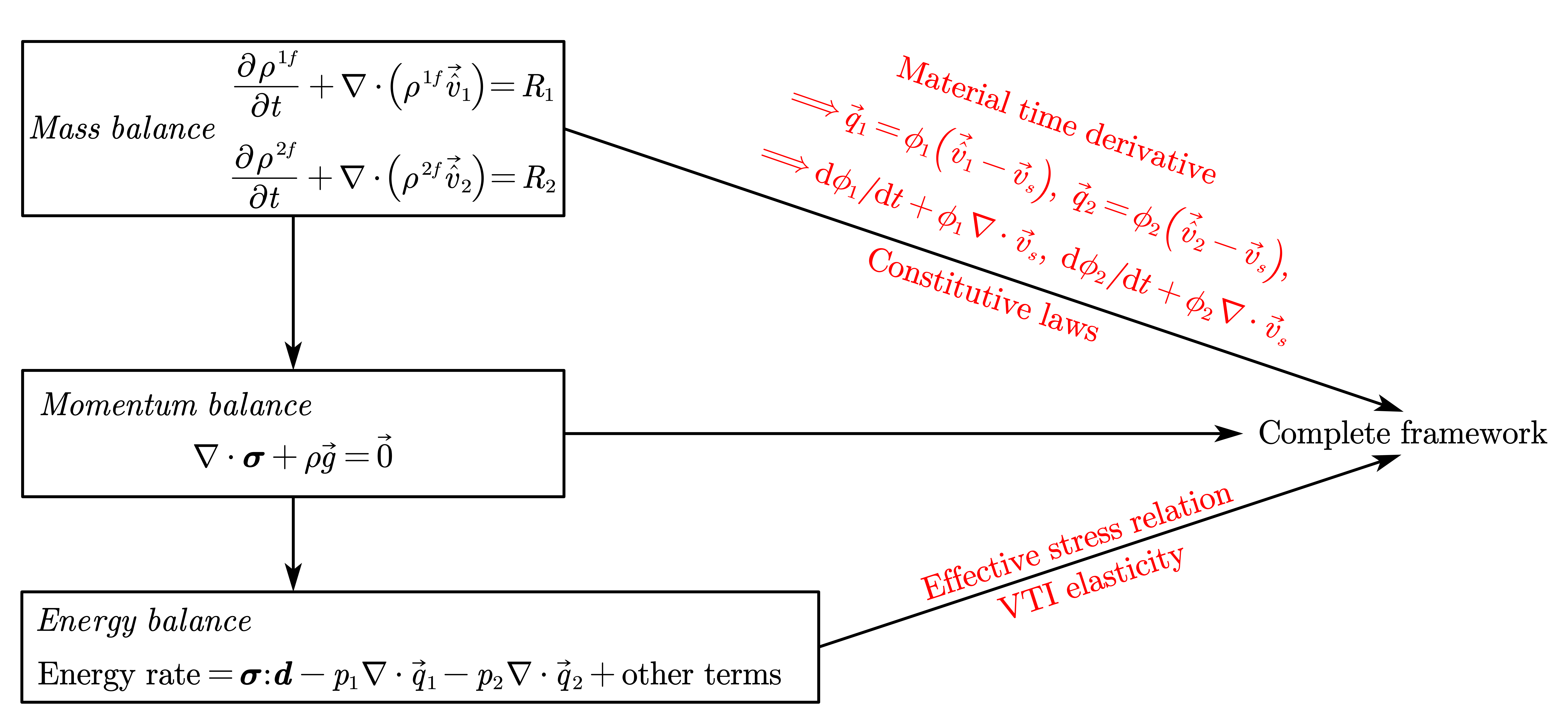}
	\end{tabular}
	\end{center}
	\caption
	{\label{Fig:03}
The interactions and relationships among different components of mathematical theories.}
\end{figure}

\textcolor{black}{The mathematical formulations given in the previous two subsections can be summarized through the following flowchart, see Figure~\ref{Fig:03}.}

\section{Computing equivalent fracture permeability}
\label{upscale}
\begin{figure}[htb]
	\begin{center}
	\begin{tabular}{c}
		\includegraphics[width=0.75\textwidth]{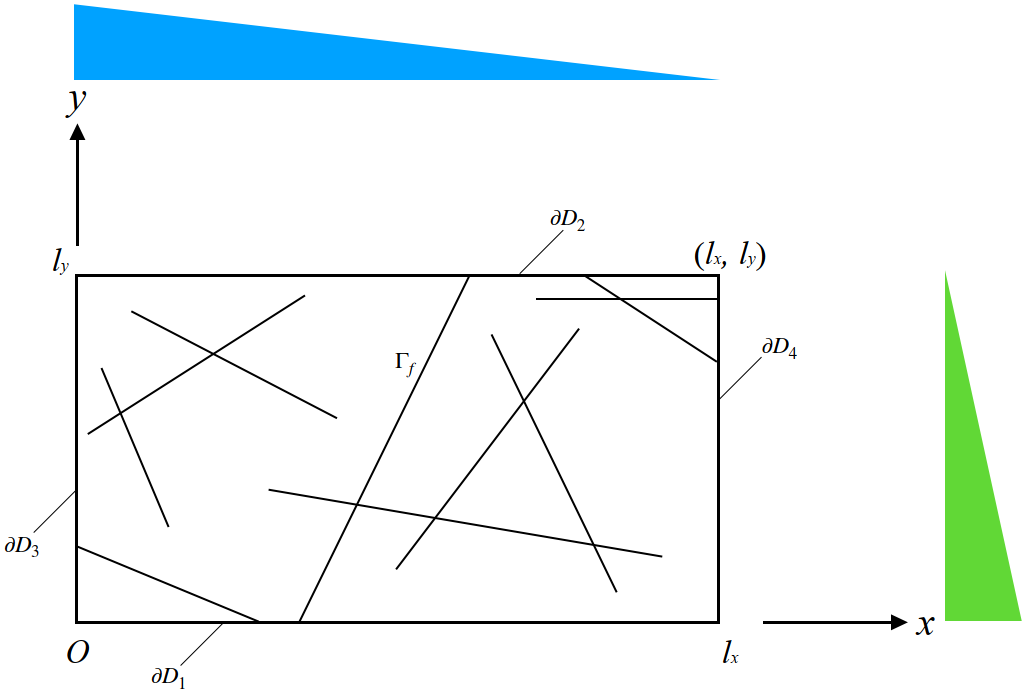}
	\end{tabular}
	\end{center}
	\caption
	{\label{Fig:05}
Schematic of a rectangular local solution domain $\Omega$ with explicit micro-fractures, the blue triangle represents the pressure boundary for the first solution, and the green triangle represents the pressure boundary for the second solution.}
\end{figure}

One of the most challenging tasks in the double porosity model is to give an accurate estimate of the equivalent permeability $\tens{k}_2$. In this section, we provide a feasible way to calculate $\tens{k}_2$. The method follows similar procedures as those described in \citet{Durlofsky2005}. Here, we illustrate this method using a purely local 2D Cartesian grid ($l_x \times l_y \times 1$) with explicit micro-fractures, see Figure~\ref{Fig:05}. We argue that extension to 3D could be done analogously. In this section, please interpret all the fractures as micro-fractures or natural fractures.

We consider two steady-state single phase flow problems. The solutions can be obtained through the finite element method by assembling the fracture flow equation into the element nodal stiffness matrix. Both solutions use linear pressure boundary conditions \citep{Durlofsky2005}. In the first solution, we set:
\begin{equation}
    p\bracs{0, y} = p_{\rm in} \qquad  y \in \partial D_3\,,
\end{equation}
\begin{equation}
    p\bracs{l_x, y} = p_{\rm out} \qquad  y \in \partial D_4\,,
\end{equation}
\begin{equation}
    p\bracs{x, 0} = p\bracs{x, l_y} = p_{\rm in} - \Delta P \frac{x}{l_x} \qquad  x \in \partial D_1 \cup \partial D_2\,,
\end{equation}
and in the second solution, we set:
\begin{equation}
    p\bracs{x, 0} = p_{\rm in} \qquad  x \in \partial D_1\,,
\end{equation}
\begin{equation}
    p\bracs{x, l_y} = p_{\rm out} \qquad  x \in \partial D_2\,,
\end{equation}
\begin{equation}
    p\bracs{0, y} = p\bracs{l_x, y} = p_{\rm in} - \Delta P \frac{y}{l_y} \qquad  y \in \partial D_3 \cup \partial D_4\,,
\end{equation}
where $p_{\rm in}$ is the inlet pressure, $p_{\rm out}$ is the outlet pressure, and $\Delta P = p_{\rm in} - p_{\rm out}$. From these two solutions, we calculate the integral of the flow velocity over the whole domain (matrix pores and micro-fractures), which can be represented as:
\begin{equation}
\label{eq:41}
    \int_{\Omega} q_i^j \d x \d y = \int_{\Omega} q_{i, \rm mat}^j \d x \d y + \sum_{\Gamma_f} e_{\Gamma_f} \int_{\Gamma_f} q_{i, \rm fr}^j \d \Gamma\,,
\end{equation}
where $\vec{q}$ is the Darcy velocity vector, the subscript $i = x, y$ represents the velocity component, the superscript $j = 1^{\rm st}, 2^{\rm nd}$ represents the solution number, $e_{\Gamma_f}$ is the aperture of the micro-fracture $\Gamma_f$. The line integral on the right-hand side of Eq.~\eqref{eq:41} represents the flow contribution from the explicit micro-fractures.

To calculate the equivalent permeability $\tens{k}_2$, we solve the same boundary value problem but modeling the explicit micro-fractures in a continuum sense. As a result, there is no inter-porosity flow, and the pressure $p_1$ and $p_2$ are given as:
\begin{equation}
    p_1 = p_2 = p_{\rm in} - \Delta P \frac{x}{l_x}
\end{equation}
for the first solution, and
\begin{equation}
    p_1 = p_2 = p_{\rm in} - \Delta P \frac{y}{l_y}
\end{equation}
for the second solution. We can again calculate the integral of the flow velocity over the whole domain, which can be represented as:
\begin{equation}
\label{eq:44}
    \int_{\Omega} \tilde{q}_i^j \d x \d y = \int_{\Omega} \tilde{q}_{i, \rm mat}^j \d x \d y + \int_{\Omega} \tilde{q}_{i, \rm fr}^j \d x \d y\,,
\end{equation}
where we add a ``tilde'' to indicate that we are dealing with double porosity media, other notations are the same as those in Eq.~\eqref{eq:41}. By comparing Eq.~\eqref{eq:44} with Eq.~\eqref{eq:41}, we could find that the line integral in Eq.~\eqref{eq:41} changes to the surface integral in Eq.~\eqref{eq:44} for micro-fractures. The $\tens{k}_2$ is incorporated into $\tilde{q}_{i, \rm fr}^j$ through Darcy's law, and the scalar matrix permeability $k_1$ can be used to calculate $\tilde{q}_{i, \rm mat}^j$ in the same manner. In this paper, we assume $k_1$ is estimated from the geometric mean of the original matrix permeabilities of Figure~\ref{Fig:05}. By equating $\int_{\Omega} \tilde{q}_i^j \d x \d y$ with $\int_{\Omega} q_i^j \d x \d y$, we can solve for $\tens{k}_2$, and the result is given as:
\begin{equation}
\label{upscale_result}
    \tens{k}_2 = 
    \begin{bmatrix}
    \bracs{\frac{\mu_f}{\Delta P l_y} \int_{\Omega} q_x^{\rm 1st} \d x \d y} - k_1 &
    \bracs{\frac{\mu_f}{2\Delta P l_y} \int_{\Omega} q_y^{\rm 1st} \d x \d y} + \bracs{\frac{\mu_f}{2\Delta P l_x} \int_{\Omega} q_x^{\rm 2nd} \d x \d y} \\
    {\rm symm} & \bracs{\frac{\mu_f}{\Delta P l_x} \int_{\Omega} q_y^{\rm 2nd} \d x \d y} - k_1
    \end{bmatrix}\,.
\end{equation}

\begin{figure}[htb]
	\begin{center}
	\begin{tabular}{c}
		\includegraphics[width=0.7\textwidth]{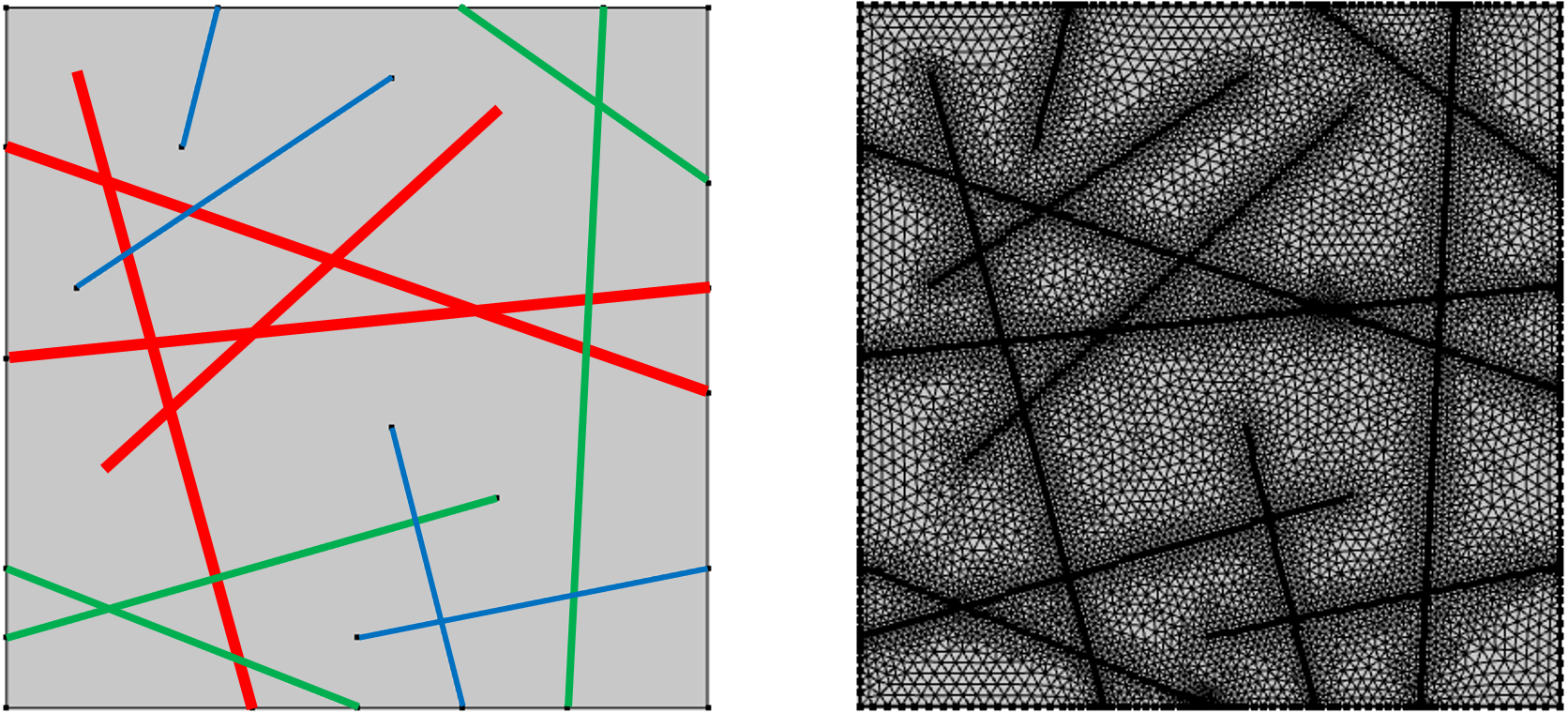}
	\end{tabular}
	\end{center}
	\caption
	{\label{Fig:06}
An actual computational domain ($1$ $\rm m$ $\times$ $1$ $\rm m$) whose scale is much smaller than the typical reservoir \citep{Yan2018,Yan2020}, and that is why the fractures drawn here can be regarded as the micro-fractures. The twelve micro-fractures could be divided into three groups based on the value of the aperture. The red group has an aperture of $5 \times 10^{-5}$ $\rm m$, the green group has an aperture of $1 \times 10^{-5}$ $\rm m$, and the blue group has an aperture of $5 \times 10^{-6}$ $\rm m$. The intrinsic fracture permeability is obtained using the cubic law. The matrix permeability $k_1$ is $10^{-18}$ $\rm m^2$, the fluid viscosity $\mu_f$ is 0.001 $\rm Pa\cdot s$, $p_{\rm in} = 1$ $\rm MPa$, and $p_{\rm out} = 0$ $\rm MPa$.}
\end{figure}

\begin{figure}[htb]
	\begin{center}
	\begin{tabular}{c}
		\includegraphics[width=0.75\textwidth]{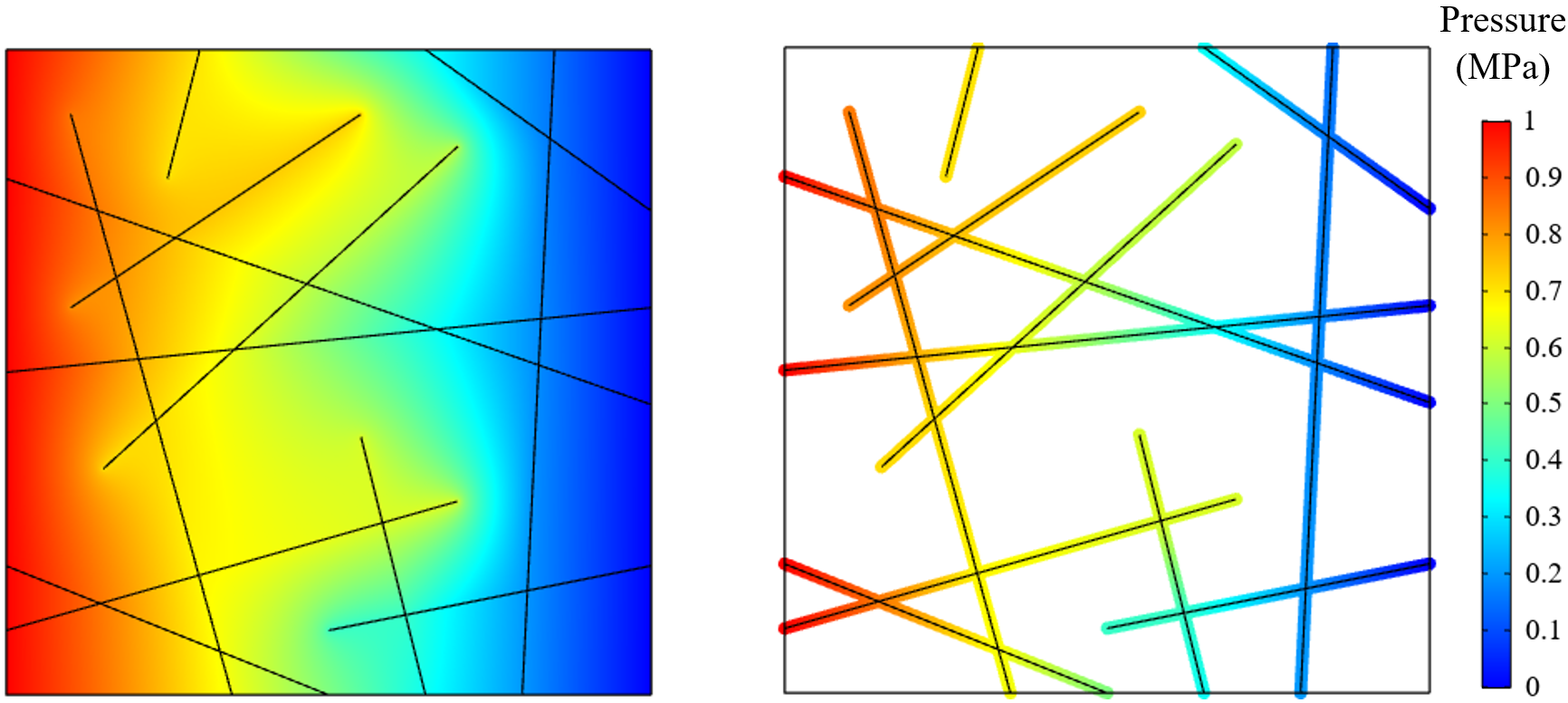}
	\end{tabular}
	\end{center}
	\caption
	{\label{Fig:07}
Pressure distribution of the first solution.}
\end{figure}

\begin{figure}[htb]
	\begin{center}
	\begin{tabular}{c}
		\includegraphics[width=0.75\textwidth]{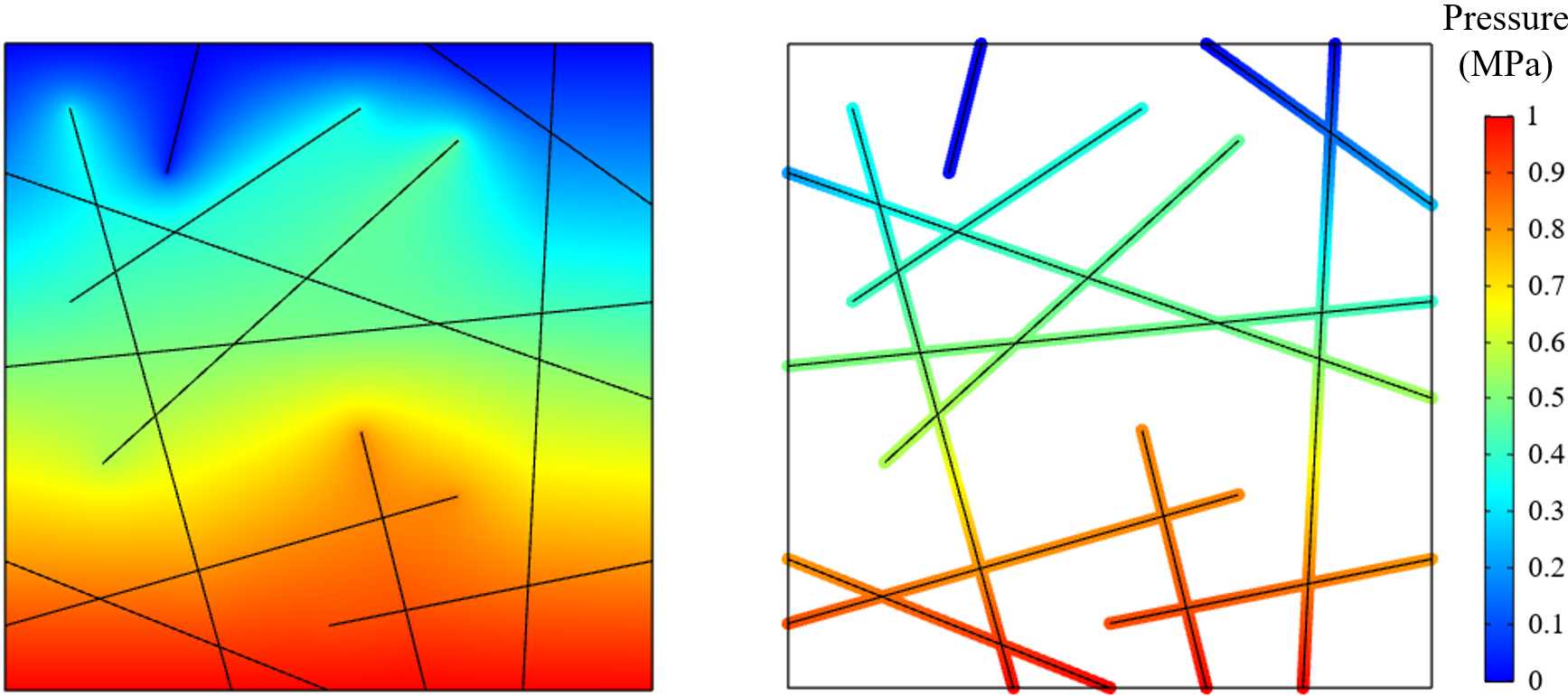}
	\end{tabular}
	\end{center}
	\caption
	{\label{Fig:08}
Pressure distribution of the second solution.}
\end{figure}

\textcolor{black}{An actual application is shown in Figure~\ref{Fig:06} to provide one numerical value of $\tens{k}_2$ using Eq.~\eqref{upscale_result}. The pressure distributions are shown in Figure~\ref{Fig:07} and Figure~\ref{Fig:08}. The equivalent $\tens{k}_2$ is calculated as:}
\begin{equation}
    \tens{k}_2 = 
    \begin{bmatrix}
    2.18\times 10^{-14} & -3.47\times 10^{-15} \\
    -3.47\times 10^{-15} & 8.68 \times 10^{-15}
    \end{bmatrix}\, {\rm m^2}\,.
\end{equation}

\begin{figure}[htb]
	\begin{center}
	\begin{tabular}{c}
		\includegraphics[width=0.7\textwidth]{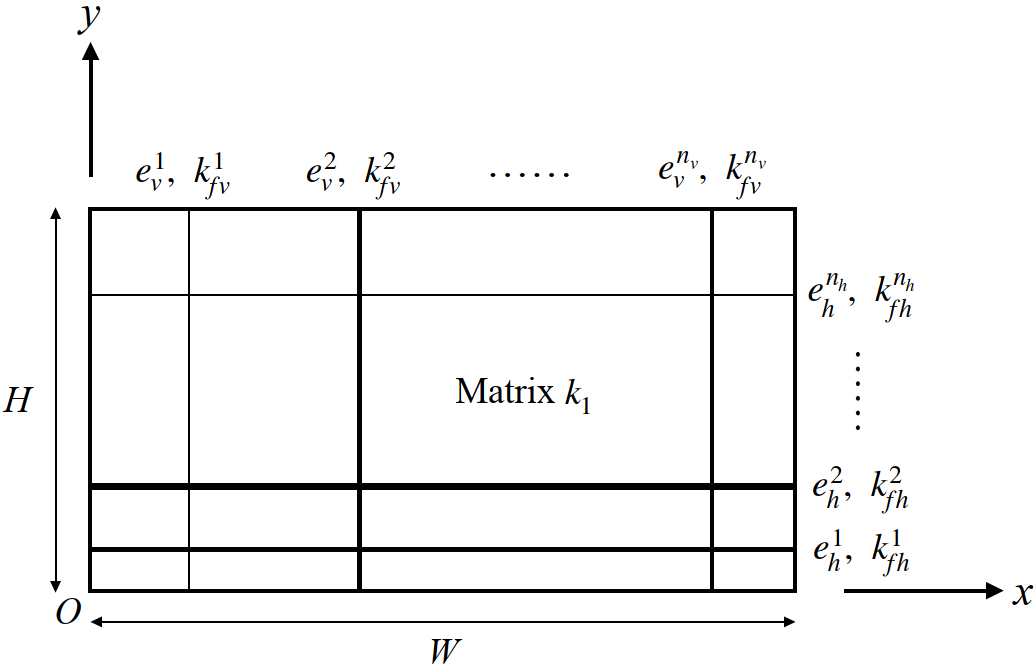}
	\end{tabular}
	\end{center}
	\caption
	{\label{Fig:09}
A special configuration allows us to calculate $\tens{k}_2$ analytically where $H$ and $W$ are used to represent the size of the domain, and other symbols are specified in the text.}
\end{figure}

In some special configurations, $\tens{k}_2$ could be obtained analytically. In Figure~\ref{Fig:09}, several horizontal and vertical micro-fractures are present within the rock matrix. These micro-fractures might have different apertures and different intrinsic fracture permeabilities. In this situation, the horizontal micro-fractures would control the $k_{2x}$, the vertical micro-fractures would control the $k_{2y}$, and the off-diagonal term $k_{2xy}$ is almost 0. The calculation process of $k_{2x}$ and $k_{2y}$ only includes the arithmetic average. The result is provided here:
\begin{equation}
    k_{2x} = \frac{1}{H}\sum_{\eta = 1}^{n_h} e_h^{\eta} k_{fh}^{\eta}\,,
\end{equation}
\begin{equation}
    k_{2y} = \frac{1}{W}\sum_{\eta = 1}^{n_v} e_v^{\eta} k_{fv}^{\eta}\,,
\end{equation}
where $e_h^{\eta}$ and $k_{fh}^{\eta}$ are the aperture and intrinsic fracture permeability of the $\eta^{\rm th}$ horizontal micro-fracture ($\eta = 1, 2, \ldots, n_h$), respectively; $e_v^{\eta}$ and $k_{fv}^{\eta}$ are the aperture and intrinsic fracture permeability of the $\eta^{\rm th}$ vertical micro-fracture ($\eta = 1, 2, \ldots, n_v$), respectively; $n_h$ and $n_v$ are the number of micro-fractures in the horizontal and vertical directions, respectively. In the following applications, we would assume a regular micro-fracture pattern and assign the value of $\tens{k}_2$ directly.

\section{Model applications}

\subsection{Consolidation of a double porosity layer}
\label{wilson}

\begin{figure}[htb]
	\begin{center}
	\begin{tabular}{c}
		\includegraphics[width=0.6\textwidth]{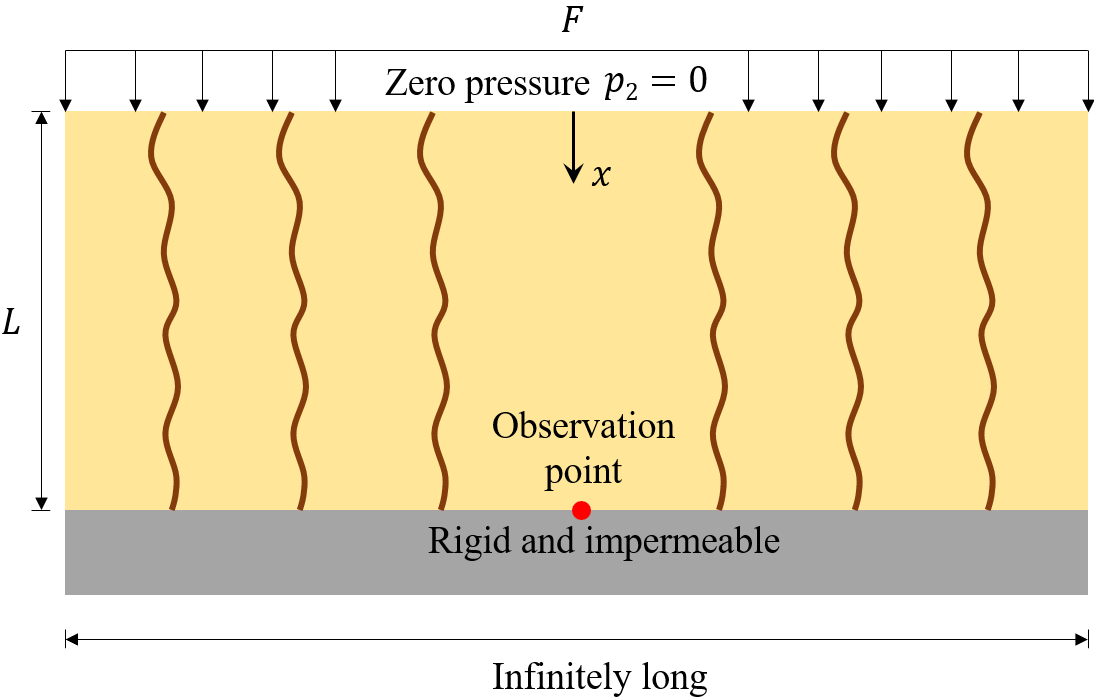}
	\end{tabular}
	\end{center}
	\caption
	{\label{Fig:10}
Conceptual illustration of the geometry and boundary conditions for the consolidation problem \citep{Ashworth2020}. Note the surface overburden $F < 0$ is not a function of time, and $L$ is the layer thickness. Gravity is not included in the analysis for this problem, \ie, excess form.}
\end{figure}

\textcolor{black}{We first apply our theory to investigate the consolidation behaviors of a double porosity layer, as shown in Figure~\ref{Fig:10}. We choose this problem since consolidation is crucial in geotechnical engineering, and by using our framework, we could have a thorough understanding of the excess pressure generation and dissipation patterns. As you can see from the following discussions, these patterns are highly dependent on the parameters you choose, and they are also significantly different from the typical pattern of Terzaghi's consolidation. Therefore, by monitoring the fluid pressure change, we can infer the degree of homogeneity and ranges of parameters for our porous medium.} Here we ignore the contribution of matrix permeability $k_1$, \ie, $\vec{q}_1 = \vec{0}$ \citep{Barenblatt1960,Warren1963,Chen1989}, and we also assume the leakage coefficient $\gamma$ is a constant. Under these assumptions, the only non-zero strain component $\epsilon_{x}$ is given as:
\begin{equation}
\label{1de_x}
    \epsilon_{x} = \frac{F + \alpha_{1x} p_1 + \alpha_{2x} p_2}{D}\,,
\end{equation}
where $D$ is the constrained modulus which depends on the values of $E_v$, $E_h$, $\nu_{vh}$, and $\nu_{hh}$. By combining Eq.~\eqref{1de_x} with Eqs.~\eqref{eq:20}\eqref{eq:21}, we could get two pressure equations:
\begin{equation}
	\label{1d_eq1}
	\bracs{A_{11} + \frac{\alpha_{1x}^2}{D}} \pd{p_1}{t} + \bracs{A_{12} + \frac{\alpha_{1x} \alpha_{2x}}{D}} \pd{p_2}{t} = \gamma \bracs{p_2 - p_1}\,,
\end{equation}
\begin{equation}
	\label{1d_eq2}
	\bracs{A_{12} + \frac{\alpha_{1x} \alpha_{2x}}{D}} \pd{p_1}{t} + \bracs{A_{22} + \frac{\alpha_{2x}^2}{D}} \pd{p_2}{t} - \frac{k_{2x}}{\mu_f} \pd[2]{p_2}{x^2} = \gamma \bracs{p_1 - p_2}\,.
\end{equation}
We also need to specify the boundary conditions for $p_2$, they are given as:
\begin{equation}
    p_2\bracs{x = 0, t > 0} = 0\,,
\end{equation}
\begin{equation}
    \left.\pd{p_2}{x}\right\vert_{x = L, t > 0} = 0\,.
\end{equation}
The initial conditions for $p_1$ and $p_2$ are $p_1\bracs{x, t = 0} = p_1^0$ and $p_2\bracs{x, t = 0} = p_2^0$, where $p_1^0$ and $p_2^0$ are obtained by solving:
\begin{equation}
	\bracs{D A_{11} + \alpha_{1x}^2} p_1^0 + \bracs{D A_{12} + \alpha_{1x} \alpha_{2x}} p_2^0 + \alpha_{1x} F = 0\,,
\end{equation}
\begin{equation}
	\bracs{D A_{12} + \alpha_{1x} \alpha_{2x}} p_1^0 + \bracs{D A_{22} + \alpha_{2x}^2} p_2^0 + \alpha_{2x} F = 0\,.
\end{equation}
We solve Eqs.~\eqref{1d_eq1}\eqref{1d_eq2} in their dimensionless forms using Laplace transform \citep{Chen1989} and numerical Laplace inversion \citep{Cheng2016,Abate2006}. In this process, we need to define following dimensionless quantities: $x_D = x/L$ is the dimensionless coordinate, $t_D = D k_{2x} t / \bracs{\mu_f L^2}$ is the dimensionless time, $p_{D1} = p_1/p_1^0$ and $p_{D2} = p_2/p_2^0$ are the dimensionless pressures, $\gamma_D = \gamma \mu_f L^2 / k_{2x}$ is the dimensionless leakage coefficient. We plot the results at $x_D = 1$, \ie, the observation point shown in Figure~\ref{Fig:10}. Besides, in order to make a better comparison, the analytical consolidation solution for single porosity media \citep{verruijt_introduction_2010, Cheng2016, Wang2000, Castelletto2015} is also included in the following figures. The solution is given as:
\begin{equation}
    p_D = \frac{p}{p^{\rm u}} = \sum_{m = 1, 3, 5, \dots}^{\infty} \frac{4}{m \pi} \sin\bracs{\frac{m \pi x_D}{2}} \exp\bracs{-\frac{m^2 \pi^2 \chi t_D}{4}}\,,
\end{equation}
where $p^{\rm u}$ is the undrained fluid pressure, and $\chi$ is a dimensionless constant which converts $t_D$ to the consistent dimensionless time $\tau = \chi t_D$ used in the typical poroelasticity analysis \citep{verruijt_introduction_2010, Cheng2016, Wang2000, Castelletto2015}.

\subsubsection{The behavior of the slightly compressible system}
\begin{figure} [htb]
	\begin{center}
	\begin{tabular}{cc}
	\includegraphics[width = 0.4\textwidth]{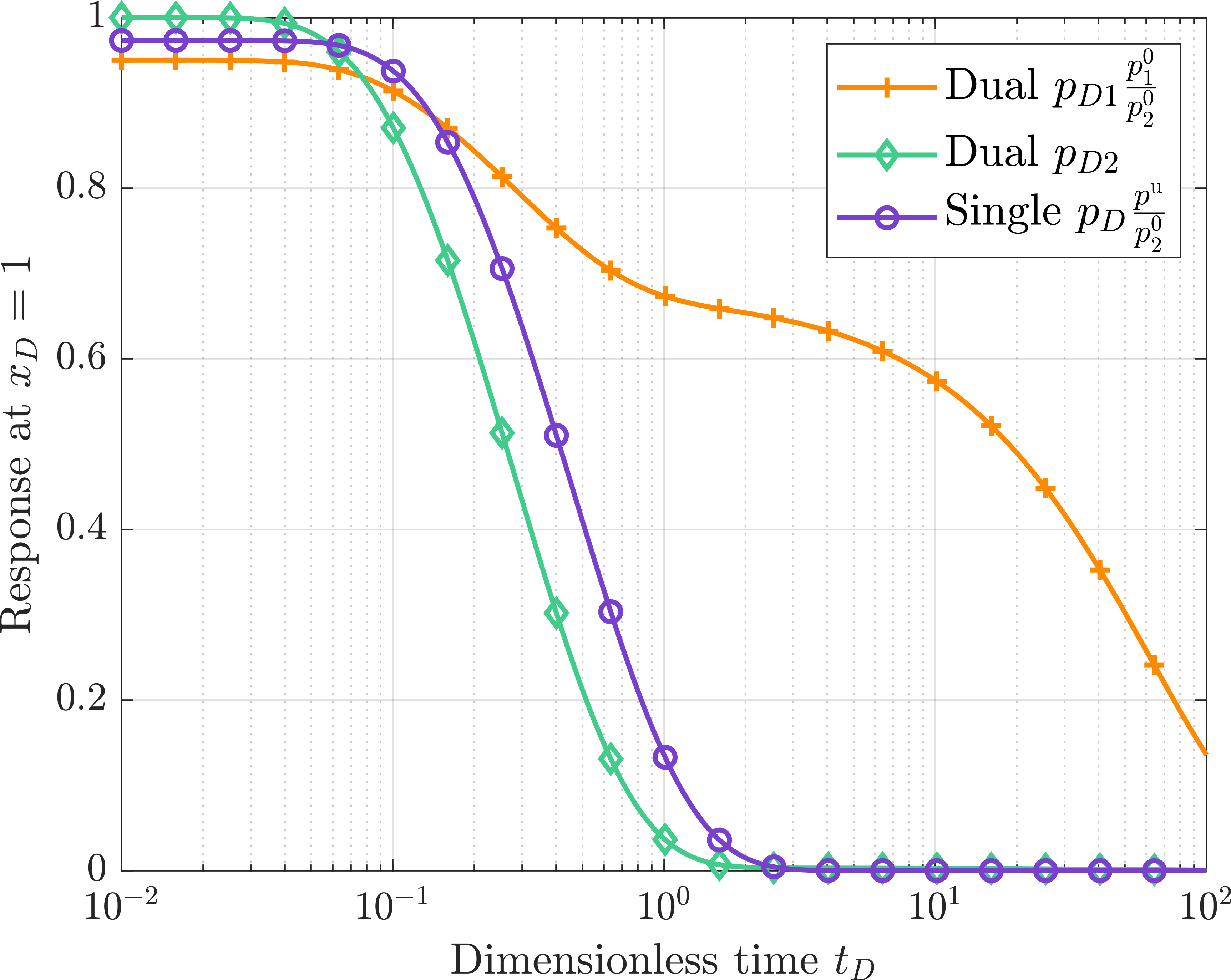} & 
	\includegraphics[width = 0.4\textwidth]{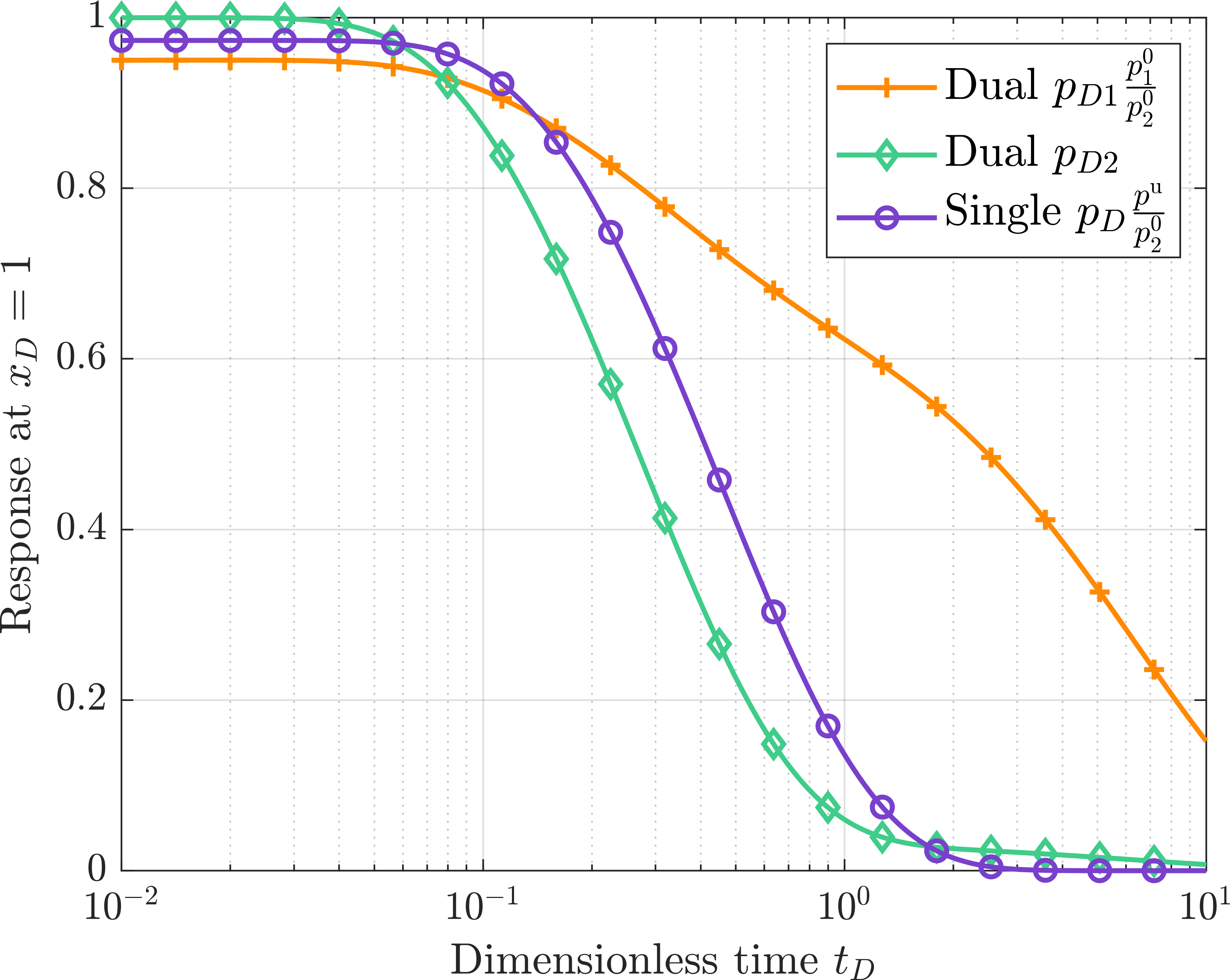} \\
	(a) $\gamma_D = 0.01336$ & (b) $\gamma_D = 0.1336$ \\
	& \\
	\includegraphics[width = 0.4\textwidth]{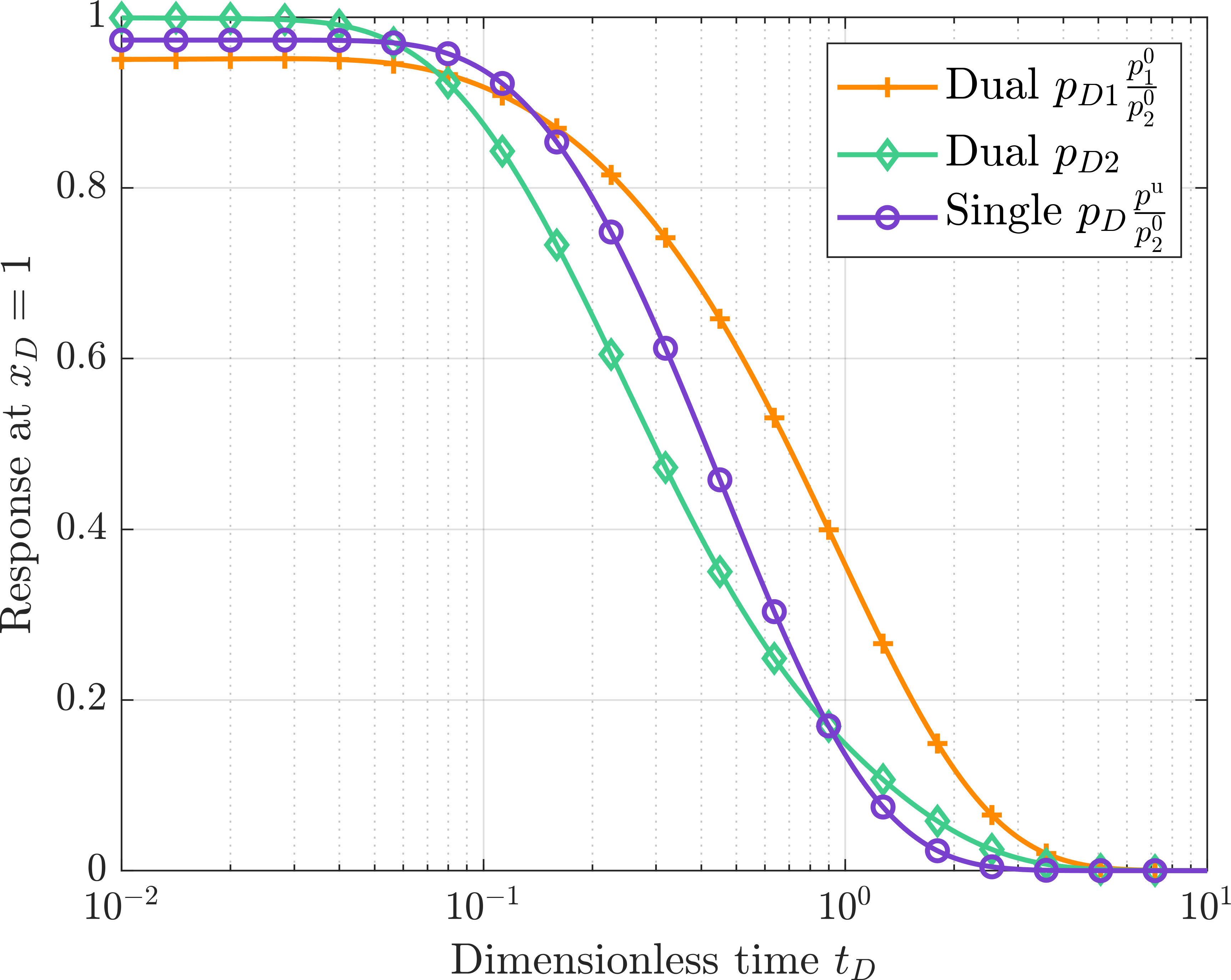} & 
	\includegraphics[width = 0.4\textwidth]{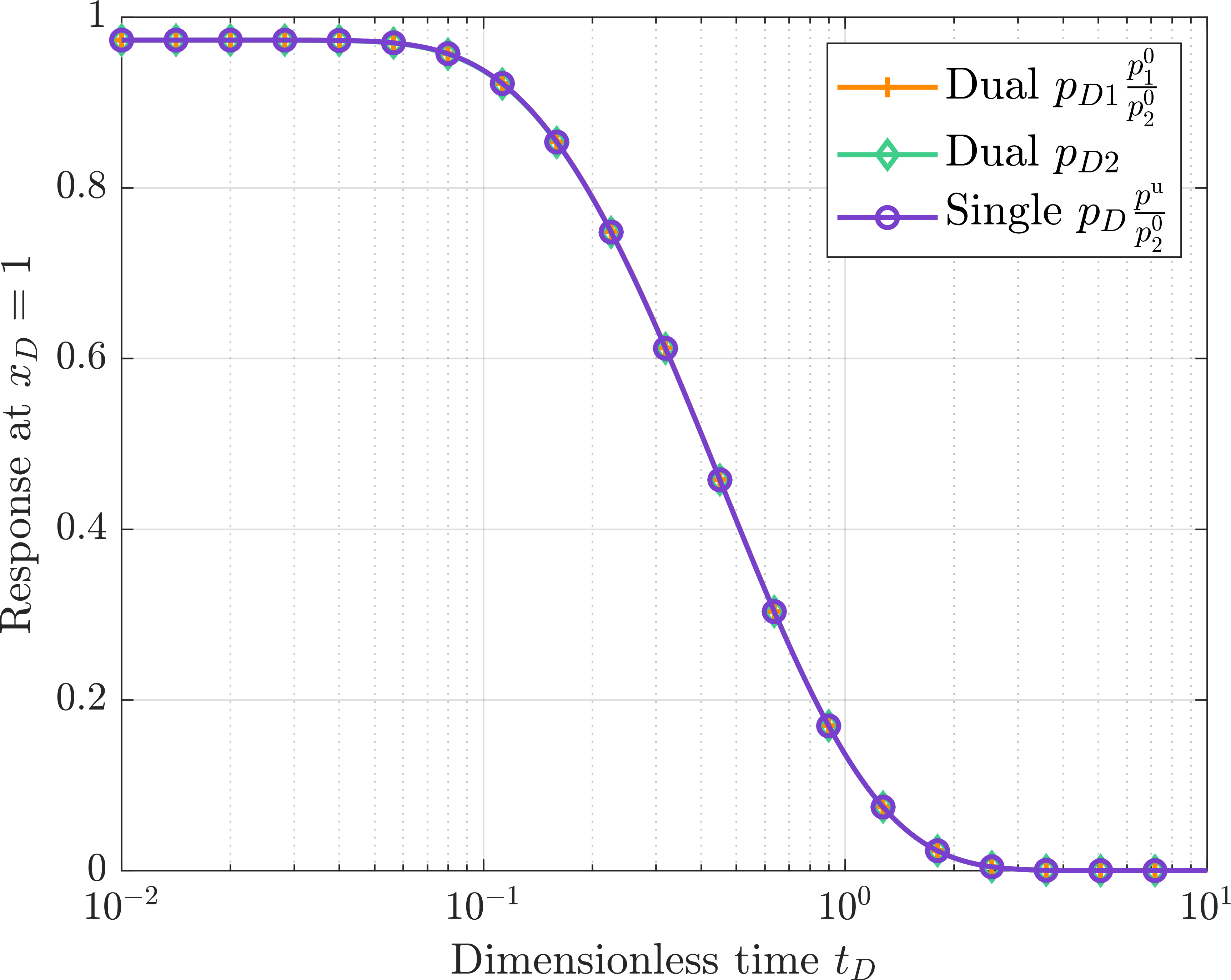} \\
	(c) $\gamma_D = 1.336$ & (d) $\gamma_D = 1336$
	\end{tabular}
	\end{center}
	\caption
	{\label{Fig:11}
Responses at $x_D = 1$ of our slightly compressible system for different $\gamma_D$.}
\end{figure}

\begin{figure} [htb]
	\begin{center}
	\begin{tabular}{c}
	\includegraphics[width = 0.6\textwidth]{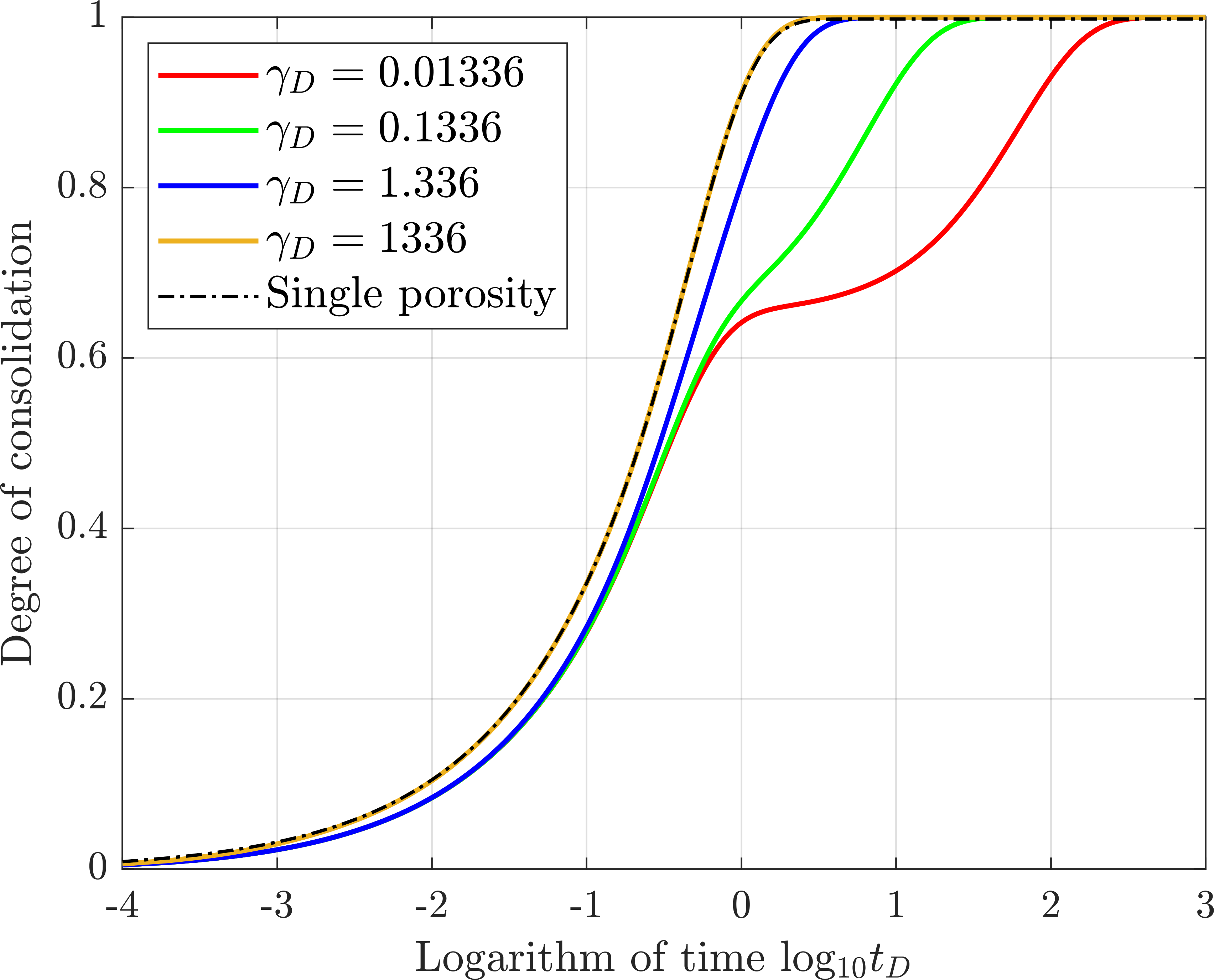}
	\end{tabular}
	\end{center}
	\caption
	{\label{Fig:11.5}
Settlement-time relationship of our slightly compressible system for different $\gamma_D$.}
\end{figure}

First of all, we consider a slightly compressible system and assume following model parameters: $D = 1.2$ $\rm GPa$, $A_{11} = 0.4825$ $\rm GPa^{-1}$, $A_{12} = -0.39289$ $\rm GPa^{-1}$, $A_{22} = 0.4357$ $\rm GPa^{-1}$, $\alpha_{1x} = 0.5007$, $\alpha_{2x} = 0.4775$. From these parameters, we could calculate $\chi = 0.8962$, $p_1^0/p_2^0 = 0.95$, and $p^{\rm u}/p_2^0 = 0.9733$. Having these values, we can plot the results under different $\gamma_D$, which are presented in Figure~\ref{Fig:11} and Figure~\ref{Fig:11.5}.

Figure~\ref{Fig:11}d suggests that for a sufficiently large $\gamma_D$, the behavior of the double porosity layer is exactly the same as that of a single porosity layer, which exhibits two stages with opposite curvatures in the logarithmic time domain. As we decrease $\gamma_D$, the pressure decline hysteresis of the double porosity layer becomes gradually obvious, as shown in Figure~\ref{Fig:11}c. In other words, the pressure decline in the transport porosity happens earlier than that of the storage porosity. If we continue decreasing $\gamma_D$, the orange curve of $p_{D1}p_1^0/p_2^0$ would have more than two distinguishable stages (known as the ``double-shell'' curve) and produce a clear separation of scales in the logarithmic time domain, which is confirmed in Figure~\ref{Fig:11}a and Figure~\ref{Fig:11}b. Furthermore, the pressure decline hysteresis is also much more evident, compared with Figure~\ref{Fig:11}c. Another finding is that in all the four cases of Figure~\ref{Fig:11}, the single porosity curve (in purple) is always caught in the middle by the double porosity curves (in orange and green), which is also true in Fig.~4 of \citet{Khalili1999}. Figure~\ref{Fig:11.5} shows that the smaller the $\gamma_D$, the longer the delay in the final settlement, and we could clearly observe two phases of consolidation (secondary compression) of the red line.

\subsubsection{The behavior of the system with a larger fluid compressibility}
\begin{figure} [htb]
	\begin{center}
	\begin{tabular}{cc}
	\includegraphics[width = 0.4\textwidth]{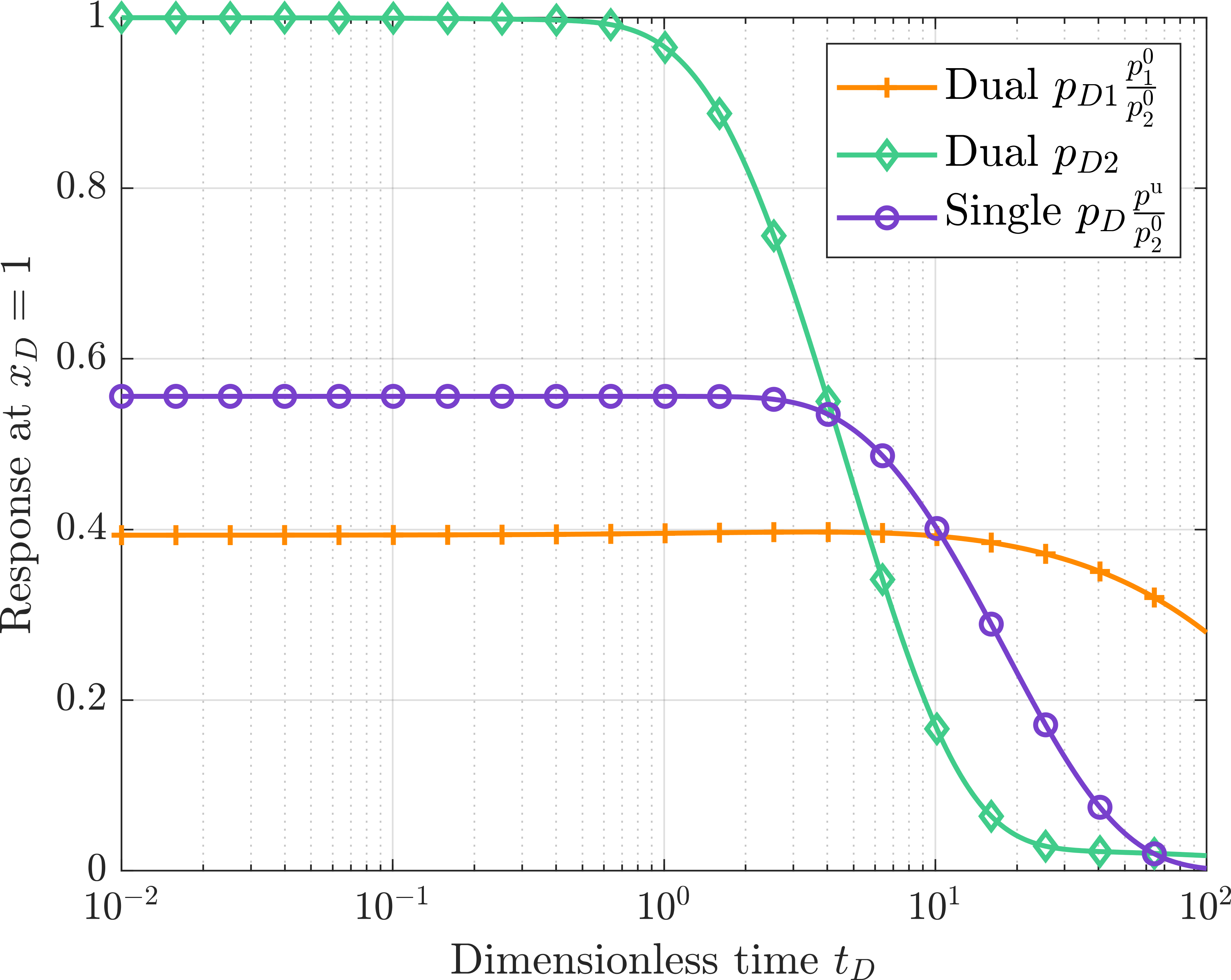} & 
	\includegraphics[width = 0.4\textwidth]{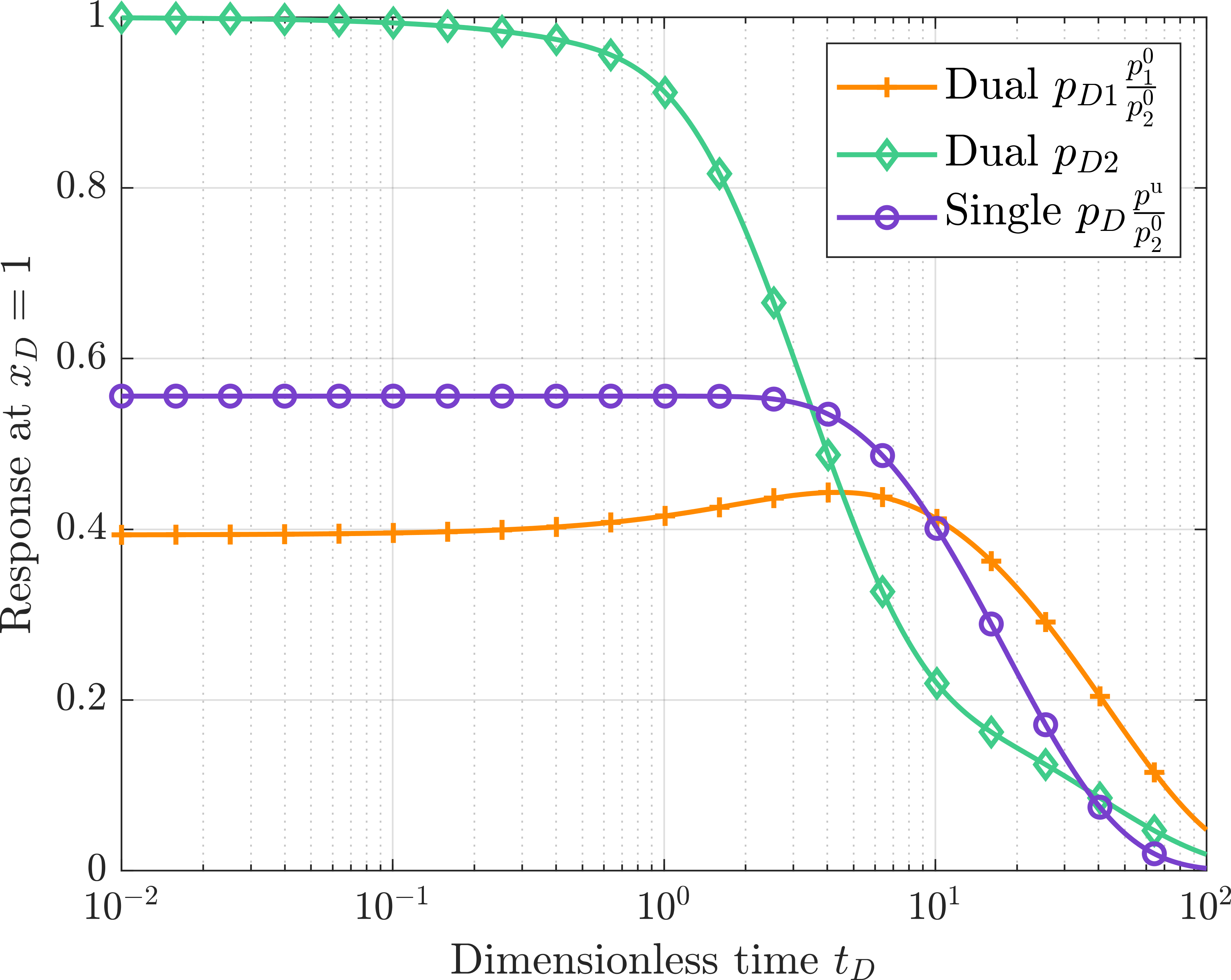} \\
	(a) $\gamma_D = 0.1336$ & (b) $\gamma_D = 1.336$ \\
	& \\
	\includegraphics[width = 0.4\textwidth]{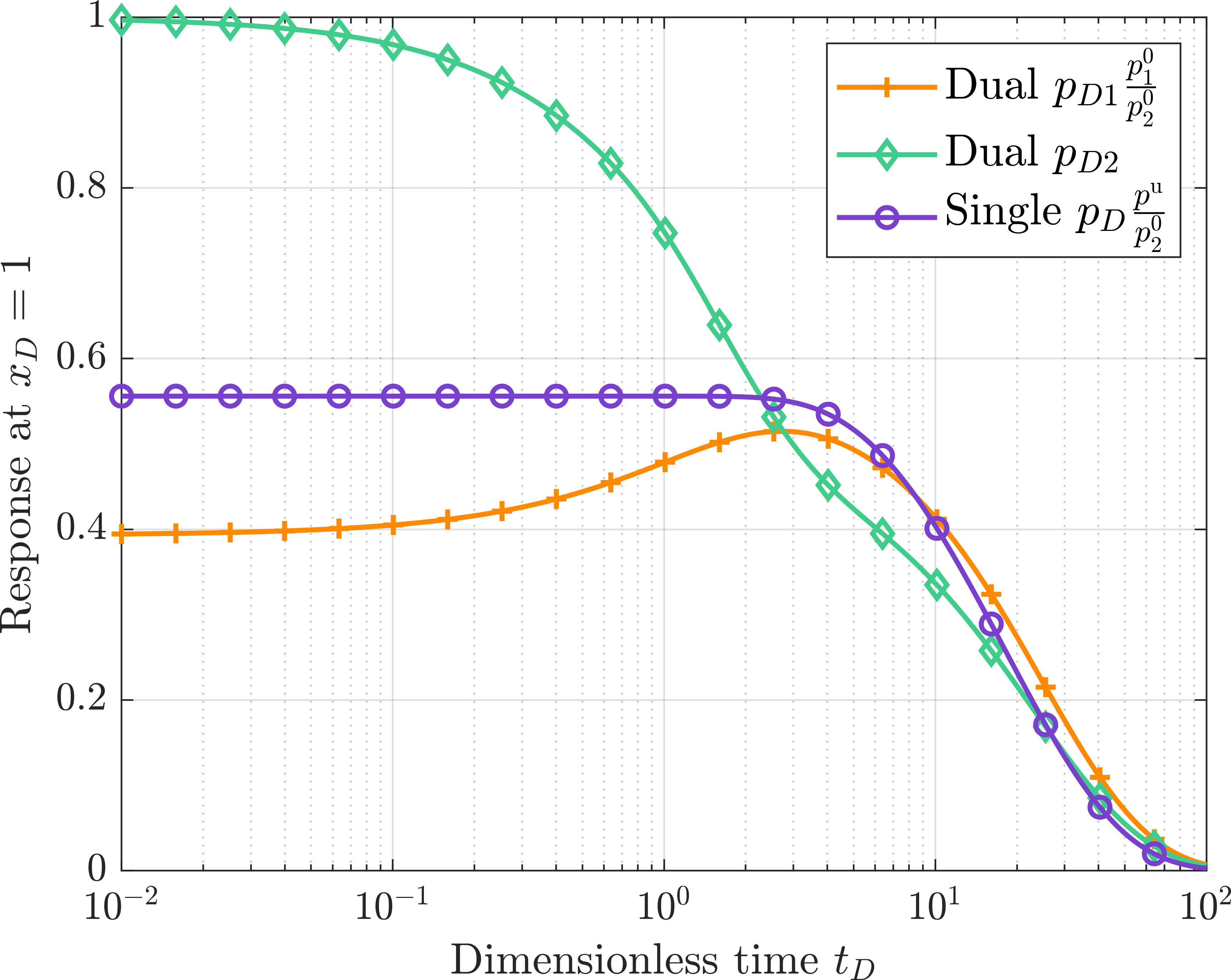} & 
	\includegraphics[width = 0.4\textwidth]{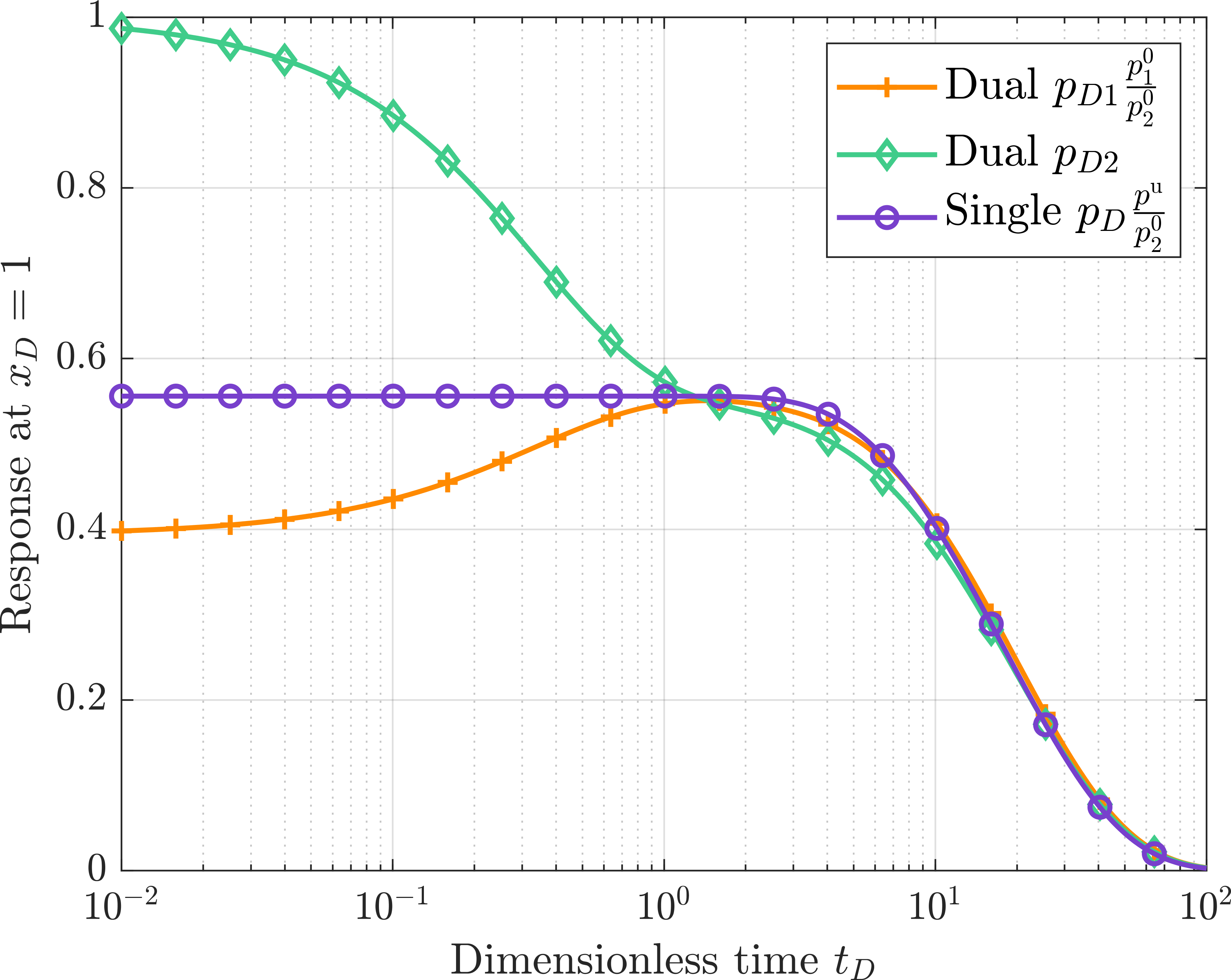} \\
	(c) $\gamma_D = 6.68$ & (d) $\gamma_D = 26.72$
	\end{tabular}
	\end{center}
	\caption
	{\label{Fig:12}
Responses at $x_D = 1$ for different $\gamma_D$ when we increase the fluid compressibility.}
\end{figure}

\begin{figure} [htb]
	\begin{center}
	\begin{tabular}{c}
	\includegraphics[width = 0.6\textwidth]{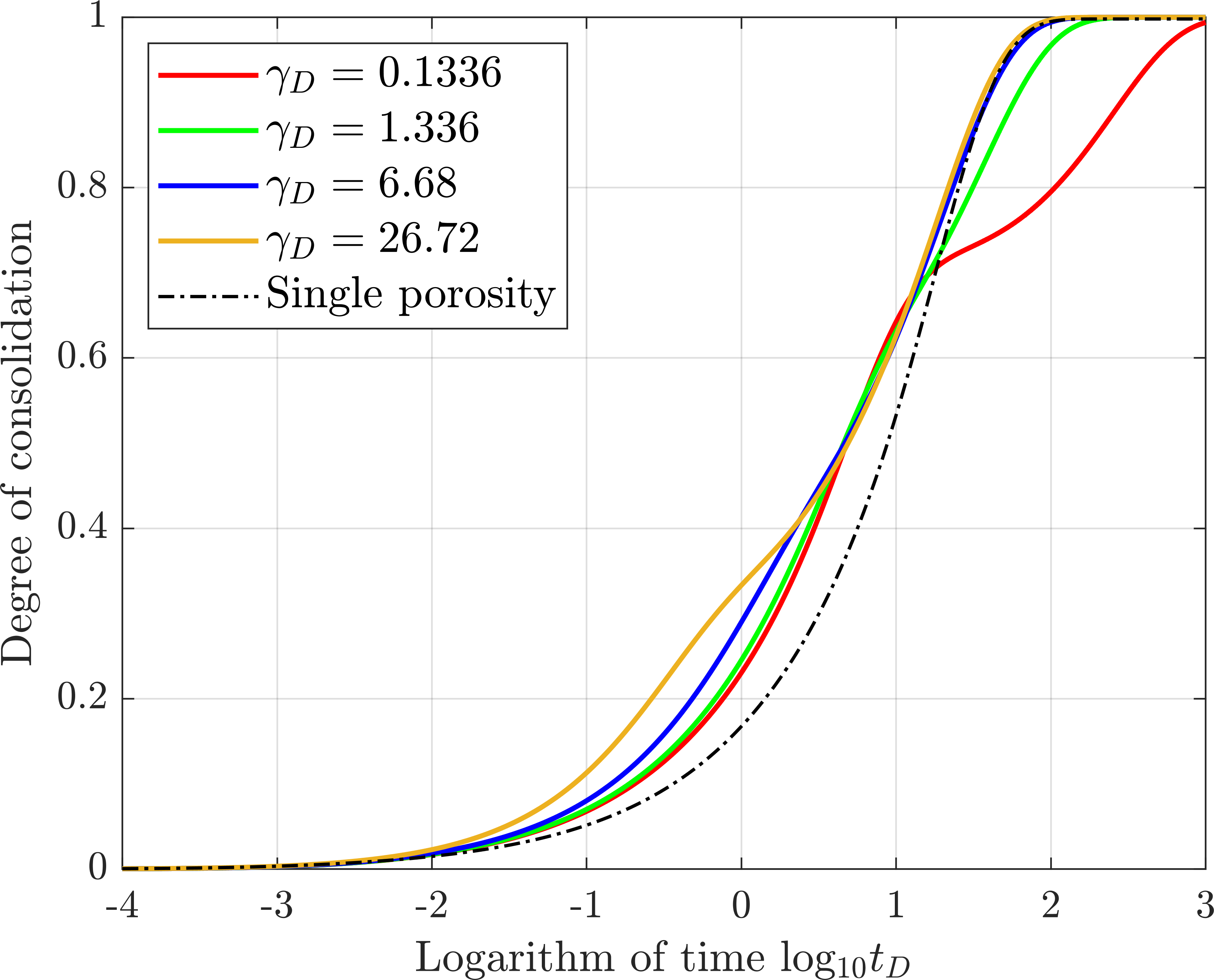}
	\end{tabular}
	\end{center}
	\caption
	{\label{Fig:12.5}
Settlement-time relationship for different $\gamma_D$ when we increase the fluid compressibility.}
\end{figure}

Now we consider the same system but with a much more compressible fluid, and as a result, some parameters need to be updated: $A_{11} = 27.0062$ $\rm GPa^{-1}$, $A_{22} = 9.9085$ $\rm GPa^{-1}$, $\chi = 0.022567$, $p_1^0/p_2^0 = 0.3933$, and $p^{\rm u}/p_2^0 = 0.556$. Several prominent features can be observed from Figure~\ref{Fig:12}. Firstly, when the fluid compressibility gets larger, the initial gap between $p_1^0$ and $p_2^0$ also gets larger, and the actual consolidation process would become longer. Secondly, when we increase $\gamma_D$, the orange curve exhibits a similar behavior as the Mandel-Cryer effect, in which the time evolution of excess pressure shows a momentary increase followed by a monotonic dissipation to zero. Also, the green curve in Figure~\ref{Fig:12}d shows the aforementioned ``double-shell'' characteristic. Compared with Figure~\ref{Fig:11.5}, Figure~\ref{Fig:12.5} shows that when we increase the fluid compressibility, the double porosity layer in the early period tends to have a larger temporary settlement than single porosity layer. These phenomena could be explained as follows: when both the $\gamma_D$ and the initial gap between $p_1^0$ and $p_2^0$ are large, the mass transfer from the transport porosity to the storage porosity would dominate the early-time response, leading to an increase in $p_{D1}$, a fast decrease in $p_{D2}$, and a large temporary settlement. Meanwhile, due to this large $\gamma_D$, we could expect an early pressure equilibrium, \ie, the behavior of the system approaches that of the single porosity layer quickly. The upshot obtained from Figure~\ref{Fig:11} to Figure~\ref{Fig:12.5} is that the responses of a double porosity layer are highly dependent on the initial pressure generation and the leakage coefficient, therefore, the ``three distinct periods'' of $p_{D2}$ proposed in \citet{Khalili1999,Khalili2003} and the ``secondary compression'' are valid only under certain ranges of parameters.

\subsubsection{The impact of matrix \texorpdfstring{$N$}{N} with incompressible solid and fluid constituents}
\label{borja}
\begin{figure} [htb]
	\begin{center}
	\begin{tabular}{cc}
	\includegraphics[width = 0.4\textwidth]{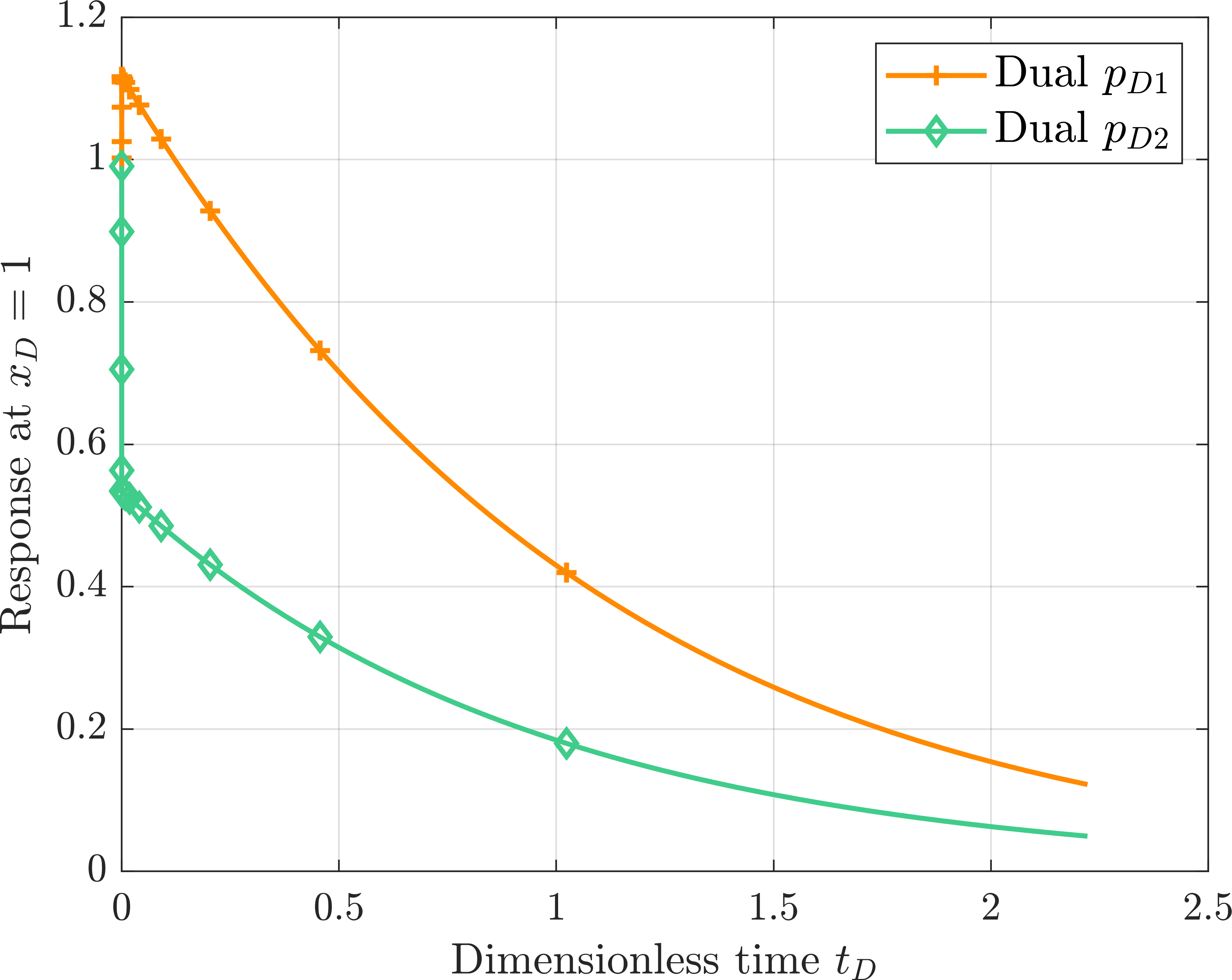} & 
	\includegraphics[width = 0.4\textwidth]{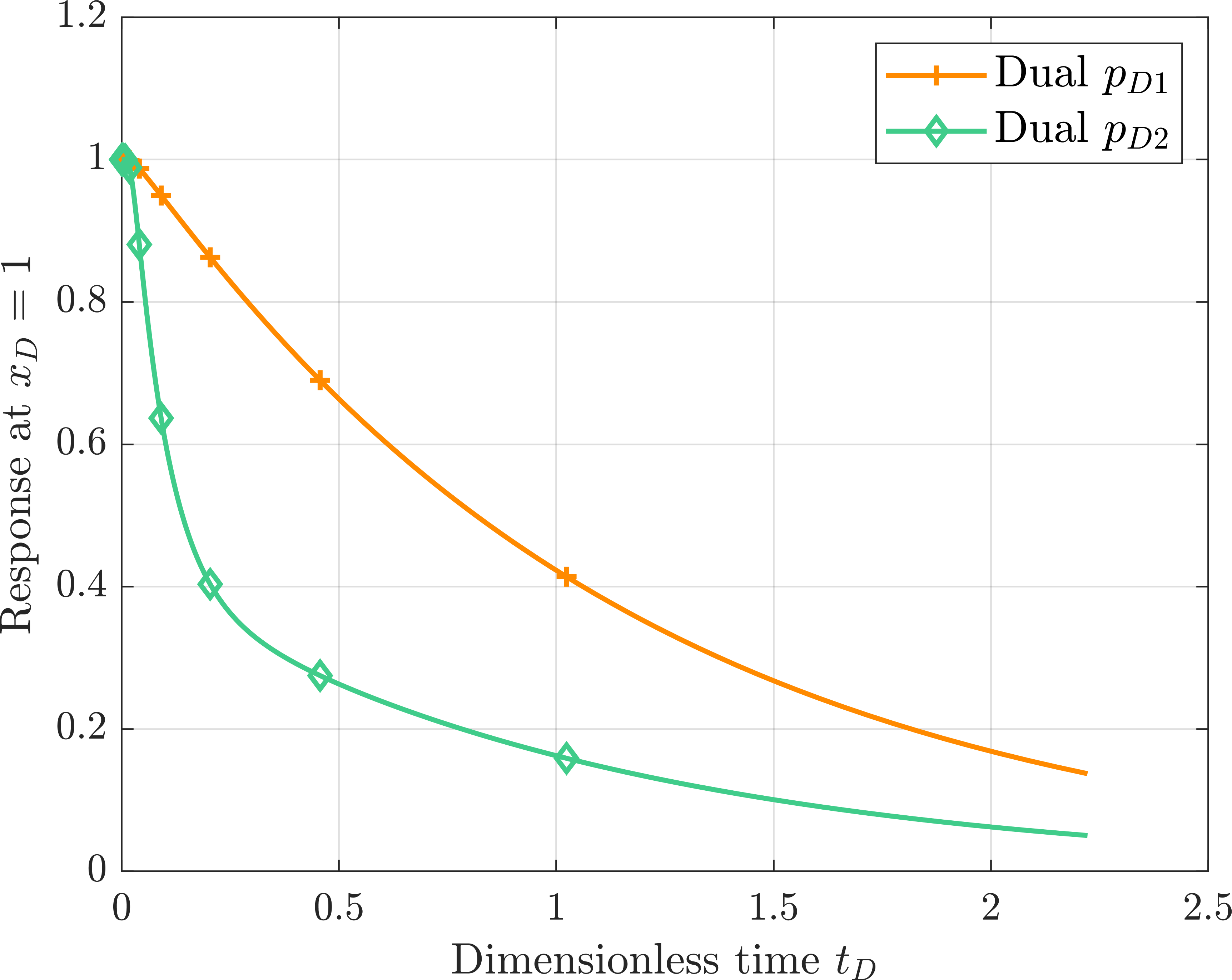} \\
	(a) $N = 0 \in \mathbb{R}^{2\times 2} $ & (b) $N \neq 0 \in \mathbb{R}^{2\times 2} $
	\end{tabular}
	\end{center}
	\caption
	{\label{Fig:13}
Responses at $x_D = 1$ for different $N \in \mathbb{R}^{2\times 2}$.}
\end{figure}

As a final note of this consolidation problem, we consider an extreme case in which $N$ is a zero matrix and $K_f \rightarrow +\infty$. This corresponds to the theory that ignores all the pressure coupling coefficients, in other words, the porosity purely depends on the volumetric strain $\epsilon_v$ (in single porosity, this is exactly the Terzaghi's assumption). To set up the problem, we assume following parameters: $K_f = 10^5$ $\rm GPa$ which is a very large number, $D = 50/9$ $\rm GPa$, $\alpha_{1x} = 0.8$, $\alpha_{2x} = 0.2$, $\phi_1 = 0.2$, $\phi_2 = 0.05$, and $\gamma_D = 1.25$. These parameters would lead to $p_1^0 = p_2^0 \approx -F$, and the calculation result is shown in Figure~\ref{Fig:13}a. From Figure~\ref{Fig:13}a, we can see that the assumption of $N = 0$ results in a discontinuous jump and a discontinuous drop in $p_{D1}$ and $p_{D2}$, respectively. This explains why in the numerical simulation, the initial pressure in the micro-fractures is always very small \citep{Zhang2020a}, and the pressure in the matrix pores always declines faster than that in the micro-fractures \citep{Zhang2020a}. As a comparison, we assume a non-zero $N$ based on \citet{Khalili1999} which is given as:
\begin{equation}
    N = \begin{bmatrix}
    0.0432 & -0.0432 \\
    -0.0432 & 0.0432
    \end{bmatrix} \, {\rm GPa^{-1}}\,.
\end{equation}
That is to say, \citet{Khalili1999} believed the $N$ is non-zero even when both the fluid and the solid constituents are incompressible. Figure~\ref{Fig:13}b displays the new result. We can find that now the pressure drop is smooth. Nevertheless, when $t_D > 1$, the difference between Figure~\ref{Fig:13}a and Figure~\ref{Fig:13}b is almost negligible, which suggests that the simplified model ($N = 0$) is more appropriate to predict the long-term behavior.

\subsection{Compression of a 3D double porosity medium}
\begin{figure} [htb]
	\begin{center}
	\begin{tabular}{c}
	\includegraphics[width = 0.55\textwidth]{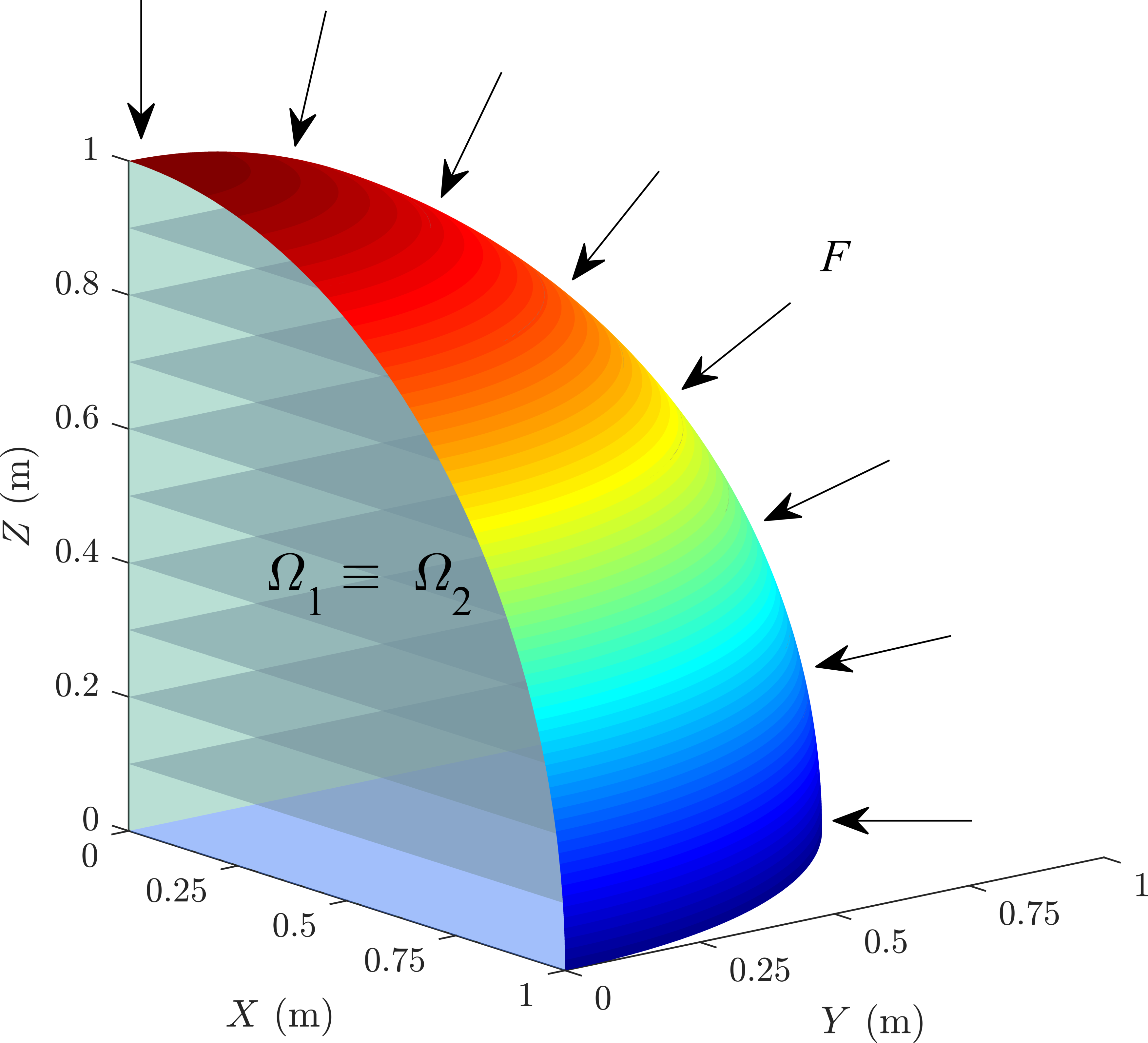}
	\end{tabular}
	\end{center}
	\caption
	{\label{numerical_modelsketch}
A sketch of the 3D compression problem. The gray horizontal planes indicate a VTI medium. Radial load ${F}$ is applied on the spherical surface with rainbow color, which also serves as a drainage boundary but only for micro-fractures. The other three surfaces that coincide with coordinate planes are impermeable with zero normal displacements. Gravity is not included in the analysis for this problem, \ie, excess form. This 3D object is meshed into $3456$ hexahedral elements.}
\end{figure}

A 3D synthetic double porosity medium is generated to contain superimposed porosity domains. The configuration of the problem and boundary (loading) conditions are shown in Figure~\ref{numerical_modelsketch}. This problem could be recast into the Cryer's problem in the limit of single porosity and isotropy \citep{White2016}. \textcolor{black}{In other words, we have enriched the poroelasticity benchmark example (3D) with double porosity.} The following model parameters are assumed in the numerical simulation: $F = 1 + 0.5\cos{X}$ $\rm MPa$, $E_v = 5000$ $\rm kPa$, $E_h = 7500$ $\rm kPa$, $\nu_{vh} = 0.25$, $\nu_{hh} = 0.15$, $G_{vh} = 3000$ $\rm kPa$, Biot coefficient ${\alpha}_1 = 0.8$,  Biot coefficient ${\alpha}_2 = 0.2$, $A_{11} = A_{12} = A_{22} = 0$, $k_1 = 10^{-15}$ $\rm m^2$ \citep{Zhao2020}, $\lambda_{\max} = 0.2306$ $\rm MPa/m$ \citep{Hao2008,Zhang2020b}, $\lambda_{\min} = 0$ \citep{Li2016}, $\xi = 2$ \citep{Li2016,Zhang2020b}, equivalent horizontal permeability $k_{2H} = 2.5 \times 10^{-13}$ $\rm m^2$, equivalent vertical permeability $k_{2V} = 5\times 10^{-15}$ $\rm m^2$, $\bar{k}_{\max} = 10^{-15}$ $\rm m^2$, $\bar{k}_{\min} = 0.75 \bar{k}_{\max}$, $\sigma_{\rm sh} = 2000$ $\rm m^{-2}$, and $\mu_f = 0.001$ $\rm Pa\cdot s$. For the simulation time configuration, we assume the initial time increment is 0.05 $\rm s$, and subsequent time increment is magnified by a factor of 1.125, \ie, $\Delta t_{n+1} = 1.125 \Delta t_n$. The total number of simulation time steps is 75.

\begin{figure}[htb]
	\begin{center}
	\begin{tabular}{c}
	\includegraphics[width = 0.8\textwidth]{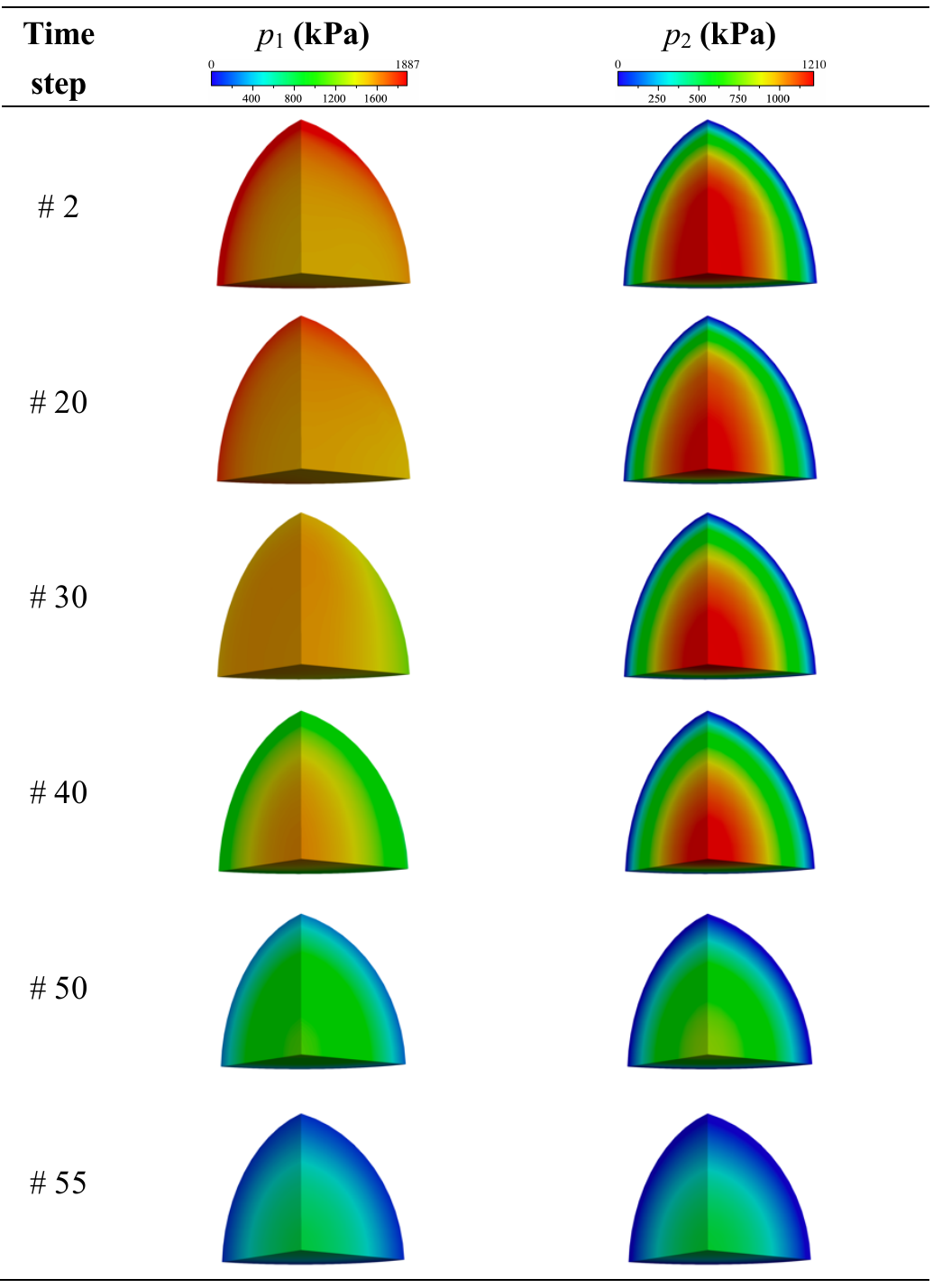}
	\end{tabular}
	\end{center}
	\caption
	{\label{pressure_surfacecontour} Evolutions of $p_1$ and $p_2$ in our deformable 3D double porosity medium.}
 \end{figure}

The evolutions of $p_1$ and $p_2$ on the deformed domain are depicted in Figure~\ref{pressure_surfacecontour}. For the first column, we can see that the early-time response of $p_1$ is dependent on the spatial distribution of $F$. For the second column, we can see that a permeability tensor with a high anisotropy ratio may not skew the pressure distribution. This is because in our case, the plane of isotropy intersects with the drainage boundary, which leads to a horizontal preferential fluid flow direction and as a result, the magnitude of $k_{2V}$ doesn't control the pattern of $p_2$. Furthermore, by comparing these two columns, we may conclude that in this case, non-equilibrium flow appears as soon as when we apply the load $F$, and it becomes weaker at a later stage, which exactly matches the observations in Figure~\ref{Fig:13}a of Section~\ref{borja}.

From the last time step, since $p_1 \approx p_2 \approx 0$, we could obtain the largest amounts of compression in all three directions, and they are $u_{x}^{\infty} = -0.071$ $\rm m$, $u_{y}^{\infty} = -0.09792$ $\rm m$, and $u_{z}^{\infty} = -0.1551$ $\rm m$. This anisotropic response is due to both the external load $F$ and the mechanical properties. In other words, $\abs{u_{x}^{\infty}} < \abs{u_{y}^{\infty}}$ is because $F$ is monotonically decreasing with $X \in \bracm{0, 1}$ $\rm m$, and $\abs{u_{y}^{\infty}} < \abs{u_{z}^{\infty}}$ is because of $E_v < E_h$.

\begin{figure}[htb]
	\begin{center}
	\begin{tabular}{c}
	\includegraphics[width = 0.6\textwidth]{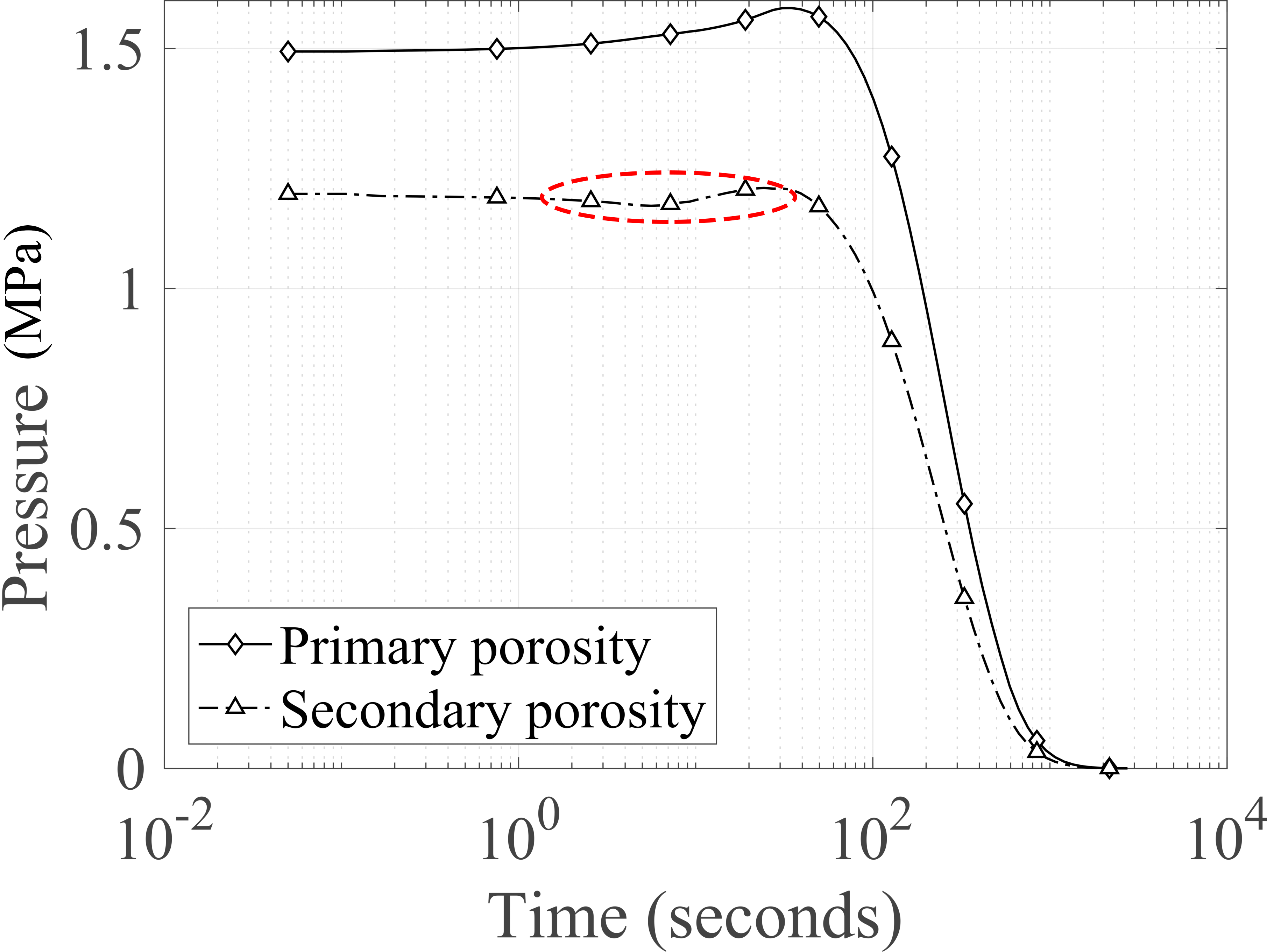}
	\end{tabular}
	\end{center}
	\caption
	{\label{last_fig} Evolutions of $p_1$ and $p_2$ at origin. The primary porosity corresponds to $p_1$ and the secondary porosity corresponds to $p_2$.}
 \end{figure}

Another interesting finding is the non-monotone characteristic of $p_2$ at origin as shown in Figure~\ref{last_fig}, which doesn't appear when $\bar{k}$ is a constant or the magnitude of $F$ is small. This difference is because when $F$ is small, the magnitude of $\vec{\varphi}$ is also small, and from Eq.~\eqref{nondarcy-leakage}, a small $\varphi_{\max}$ leads to $\bar{k} \equiv \bar{k}_{\min}$, so we would expect the same behaviors as when $\bar{k}$ is a constant. That's to say, the non-monotone characteristic of $p_2$ is related to a changing $\bar{k}$ and a broad range of $\varphi_{\max}$ compared to $\lambda_{\max}$. Note the non-monotone characteristic of $p_1$ is a natural result of load transfer from the drainage boundary to the origin \citep{Wang2000}.

\section{Closure}
We have presented a comprehensive continuum framework for anisotropic and deformable porous materials exhibiting two dominant porosity scales with ultra-low matrix permeability. Through mathematical formulations, we have identified challenges in modeling $\mathrm{d}\phi_1/\mathrm{d}t$ and $\mathrm{d}\phi_2/\mathrm{d}t$. As a result, anisotropic constitutive equations for the porosities changes are proposed for the first time. An upscaling approach based on the volume integral is also proposed in this work to fill the gap between discrete geological descriptions and equivalent fracture permeability $\tens{k}_2$. \textcolor{black}{Finally, for the model applications, we have used this framework to give a thorough discussion of consolidation with double porosity, in which we have discovered many unique patterns that were not reported in the previous publications. We have also enriched the poroelasticity benchmark example (3D) with double porosity. We believe in the future, our framework could be combined with some discrete fracture methods to form a hybrid model, which will be an excellent candidate to simulate the shale gas reservoir after hydraulic fracturing.}

\section*{CRediT authorship contribution statement}
\textbf{Qi Zhang:} Conceptualization, Methodology, Software, Writing - Original Draft, Writing - Review \& Editing. \textbf{Xia Yan:} Validation, Formal analysis. \textbf{Jianli Shao:} Visualization, Writing - Review \& Editing.

\section*{Declaration of Competing Interest}
The authors declare that they have no known competing financial interests or personal relationships that could have appeared to influence the work reported in this paper.

\section*{Acknowledgments}

% This research was funded by the National Natural Science Foundation of China (51774199, \ 52004321), the Natural Science Foundation of Shandong Province for the support of major basic research projects (ZR2018ZC0740), the China Postdoctoral Science Foundation (2020M682265), the Postdoctoral Innovation Fund of Shandong Province (202003016), the Fundamental Research Funds for the Central Universities (20CX06025A), and the Qingdao Postdoctoral Applied Research Project (QDYY20190025). The first author acknowledges financial support provided by the John A. Blume Earthquake Engineering Center at Stanford University. The authors are grateful to the anonymous reviewer for the constructive comments. We also deeply thank Prof.~Ronaldo I. Borja (Stanford University) and Mr.~Mark Ashworth (Heriot-Watt University) for comments and suggestions on the manuscript. In particular, Prof.~Ronaldo I. Borja's expertise has helped to improve the manuscript substantially.

\bibliographystyle{cas-model2-names}
\bibliography{refbib}

\end{document}